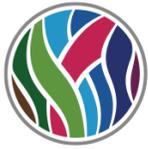

## IN PRESS

# The rise and fall of the *Phytophthora infestans* lineage that triggered the Irish potato famine

The late blight pandemic that included the Irish Great Famine in the nineteenth century was caused by a single *Phytohpthora infestans* genotype, which is distinct but closely related to the most prevalent genotype of the twentieth century.

Kentaro Yoshida (The Sainsbury Laboratory), Verena Schuenemann (University of Tübingen), Liliana Cano (The Sainsbury Laboratory), Marina Pais (The Sainsbury Laboratory), Bagdevi Mishra (Biodiversity and Climate Research Centre), Rahul Sharma (Biodiversity and Climate Research Centre), Christa Lanz (Max Planck Institute for Developmental Biology), Frank Martin (United States Department of Agriculture), Sophien Kamoun (The Sainsbury Laboratory), Johannes Krause (University of Tübingen), Marco Thines (Biodiversity and Climate Research Centre), Detlef Weigel (Max Planck Institute for Developmental Biology), and Hernán Burbano (Max Planck Institute for Developmental Biology)

**Abstract:**

*Phytophthora infestans*, the cause of potato late blight, is infamous for having triggered the Irish Great Famine in the 1840s. Until the late 1970s, *P. infestans* diversity outside of its Mexican center of origin was low, and one scenario held that a single strain, US-1, had dominated the global population for 150 years; this was later challenged based on DNA analysis of historical herbarium specimens. We have compared the genomes of 11 herbarium and 15 modern strains. We conclude that the nineteenth century epidemic was caused by a unique genotype, HERB-1, that persisted for over 50 years. HERB-1 is distinct from all examined modern strains, but it is a close relative of US-1, which replaced it outside of Mexico in the twentieth century. We propose that HERB-1 and US-1 emerged from a metapopulation that was established in the early 1800s outside of the species' center of diversity.

http://dx.doi.org/10.7554/elife.00731

34    **Introduction**

35    Potato late blight's impact on humankind is rivaled by few other plant diseases. The Spanish
36    introduced Europeans to the South American staple crop potato shortly after their conquest of
37    the New World, but for three centuries Europe stayed free of *P. infestans*, the causal agent of
38    late blight. In 1845, the oomycete *P. infestans* finally reached Europe, spreading rapidly from
39    Belgium to other countries of mainland Europe and then to Great Britain and Ireland. The
40    impact of the epidemic reached catastrophic levels in Ireland, where the population was more
41    dependent on potato for their subsistence than in other parts of Europe (Bourke, 1964;
42    Reader, 2009). The subsequent Great Famine killed around one million people, and an
43    additional million were forced to leave the island (Turner, 2005). Even today, the Irish
44    population remains less than three quarters of what it was at the beginning of the 1840s.
45    These dramatic consequences of the *P. infestans* epidemic were due to the absence of
46    chemical and genetic methods to combat it; such means became available only several
47    decades later.

48        Ever since triggering the Irish famine, *P. infestans* has continued to wreak havoc on
49    potato fields throughout the world. Late blight remains the most destructive disease of the
50    third largest food crop, resulting in annual losses of potatoes that would be sufficient to feed
51    anywhere from 80 to many hundreds of millions of people (Fisher et al., 2012). *Phytophthora*
52    *infestans* is an extraordinarily virulent and adaptable pathogen (Fry, 2008; Haas et al., 2009).
53    In agricultural systems, sexual reproduction may trigger explosive population shifts that are
54    driven by the emergence and migration of asexual lineages (Cooke et al., 2012; Fry et al.,
55    2009; Fry et al., 1992). The species is thought to originate from Toluca Valley, Mexico,
56    where it infects wild relatives of potato, frequently undergoes sexual reproduction and co-
57    occurs with the two closely related species *P. mirabilis* and *P. ipomoeae* (Flier et al., 2003;
58    Goodwin et al., 1994; Grünwald and Flier, 2005; Tooley et al., 1985). In its center of origin,
59    *P. infestans* is characterized by high levels of genetic and phenotypic diversity (Grünwald and
60    Flier, 2005).

61        The genomes of a few *P. infestans* strains have been described (Cooke et al., 2012;
62    Haas et al., 2009; Raffaele et al., 2010a). Compared to other species in the genus, the 240 Mb
63    T30-4 reference genome of *P. infestans* is large, with three quarters of the genome consisting
64    of repetitive DNA. A large number of genes codes for effector proteins, many of which are
65    delivered inside plant cells to promote host colonization, for instance by suppressing plant
66    immunity. RXLR proteins, the main class of host-translocated effectors, are encoded by about

67  550 genes in the *P. infestans* T30-4 genome. RXLR effectors that can be recognized by plant
68  immune receptors, known as Resistance (R) proteins, are said to have "avirulence" activity.
69  Upon introduction of a cognate *R* gene into the host population, such avirulence effectors
70  become a liability for the pathogen, and natural selection favors the spread of pseudogenized
71  or mutated alleles (Vleeshouwers et al., 2011).

72      The detailed descriptions and drawings of Heinrich Anton de Bary and the reports of
73  several other pioneers of plant pathology leave no doubt that the nineteenth century blight
74  epidemic was triggered by *P. infestans* (Bourke, 1964; de Bary, 1876). What remains
75  controversial is the relationship of the nineteenth century strains to modern isolates. The quest
76  for understanding the origin of the strain that resulted in the Irish famine began with extant
77  samples. Prior to the late 1970s, global *P. infestans* populations outside of South America and
78  Mexico, the centers of diversity of the host and the pathogen, were dominated by a single
79  clonal lineage that had the mitochondrial (mtDNA) haplotype Ib and was called US-1
80  (Goodwin et al., 1994). It was therefore proposed that the US-1 lineage was a direct
81  descendant of the strain that first caused widespread late blight in North America from 1843
82  on, and then triggered the Irish famine beginning in 1845 (Bourke, 1964; Goodwin et al.,
83  1994). This hypothesis was subsequently directly addressed by PCR analysis of infected
84  nineteenth century potato leaves stored in herbaria. The conclusion from these studies was
85  that the historic strains belonged to a mtDNA haplotype, Ia, that was distinct from that of the
86  US-1 lineage (May and Ristaino, 2004; Ristaino et al., 2001). Because Ia was at the time not
87  only the predominant haplotype in the Toluca Valley in Mexico (Flier et al., 2003; Gavino
88  and Fry, 2002), but had also been found in South America (Perez et al., 2001), May and
89  Ristaino (2004) speculated that the nineteenth century and US-1 lineages represented two
90  independent epidemics of divergent lineages that had both originated in South America and
91  spread from there to North America and Europe. A caveat was that these far-reaching
92  conclusions were based on only three mtDNA SNPs (May and Ristaino, 2004; Ristaino et al.,
93  2001).

94      Since these first herbarium analyses, the retrieval and sequencing of DNA from
95  museum specimens, fossil remains and archaeological samples – collectively known as
96  ancient DNA (aDNA) (Pääbo et al., 2004) – have seen impressive advances thanks to the
97  advent of high-throughput sequencing technologies. The combined analysis of modern and
98  ancient genomes of human pathogens has begun to solve important questions about their
99  history and evolution (Bos et al., 2011; Bos et al., 2012). Here we show that aDNA methods

100   hold similar promise for plant pathology and that they can improve our understanding of
101   historically important plant pathogen epidemics.

102      To determine how the historic *P. infestans* strain(s) relate to extant isolates, we
103   shotgun-sequenced 11 herbarium samples of infected potato and tomato leaves collected from
104   continental Europe, Great Britain, Ireland, and North America in the period from 1845 to
105   1896, and extracted information on *P. infestans* mitochondrial and nuclear genomes. To
106   understand the subsequent evolution of the pathogen, we compared the historic *P. infestans*
107   genomes to those of 15 modern twentieth century strains that span the genetic diversity of the
108   species, and to the two sister species *P. ipomoeae* and *P. mirabilis*. Our analyses revealed that
109   the nineteenth century epidemic was caused by a single genotype, HERB-1, that persisted for
110   at least 50 years. While it is distinct from all examined modern strains, HERB-1 is closely
111   related to the twentieth century US-1 genotype, suggesting that these two pandemic genotypes
112   may have emerged from a secondary metapopulation rather than from the species' Mexican
113   center of diversity.

## Results

### Preservation of ancient DNA and genome statistics

116   Nineteenth-century samples of potato and tomato leaves with *P. infestans* lesions were
117   obtained from the herbaria of the Botanische Staatssammlung München and the Kew Royal
118   Botanical Gardens (Table 1 and Figure 1). DNA was extracted under clean room conditions
119   and two genomic libraries were prepared from each sample for Illumina sequencing. The
120   preparations were expected to comprise *P. infestans* DNA, host DNA from potato or tomato
121   as well as DNA from microbes that had colonized either the living material at the time of its
122   collection, or the dried material during its storage in the herbaria.

123      The first set of libraries was used for verification of aDNA-like characteristics, and the
124   second set was used for production. In this second set we used a repair protocol that removes
125   aDNA-associated lesions, mainly characterized by cytosine deamination to uracil (U), which
126   is subsequently converted to and read as thymine (T) (Briggs et al., 2007; Briggs et al., 2010;
127   Brotherton et al., 2007; Hofreiter et al., 2001). The combination of uracil-DNA-glycosylase
128   (UDG) and endonuclease VIII, which removes uracil residues and repairs abasic sites, reduces
129   the overall per-base error rate to as low as one twentieth of unrepaired aDNA (Briggs et al.,
130   2010).

131     Ancient DNA fragments are typically shorter than 100 bp (Pääbo, 1989), and paired-
132     end reads of 100 bases each will therefore substantially overlap. Forward and reverse reads
133     from the unrepaired libraries (Table 2) were merged, requiring at least 11 base overlap.
134     Merging of short-insert libraries considerably decreases the error-rate and also generates
135     sequences that reflect the original molecule length (Kircher, 2012). The median length of
136     merged reads was in the range of ~50-85 bp (Figure 2a, b).

137     Merged sequences were mapped to the *P. infestans* T30-4 reference genome (Haas et
138     al., 2009). Deamination of C to U in aDNA is highest at the first base (Briggs et al., 2007),
139     and C-to-T was the predominant substitution at the 5'-end of molecules (Figure 2c, d). Based
140     on mapping against the reference genome, we estimated the fraction of *P. infestans* DNA in
141     the samples to be between 1 and 20% (Figure 2e). Most of the remaining reads could be
142     mapped to the reference genomes for potato and tomato
143     (Potato Genome Sequencing Consortium, 2011; The Tomato Genome Consortium, 2012).

144     In addition to 11 historic samples from Ireland, Great Britain, continental Europe and
145     North America (Figure 1 and Table 1), we shotgun sequenced 14 modern strains from
146     Europe, the Americas and Africa (Figure 1 and Table 1). These had been selected based on
147     preliminary mtDNA information to present a cross section of *P. infestans* diversity. Finally,
148     we sequenced two strains of *P. mirabilis*, P7722 and PIC99114, and a single strain of *P.*
149     *ipomoeae*, PIC99167. Both species are closely related to *P. infestans* and served as outgroups
150     (Kroon et al., 2004; Raffaele et al., 2010a). We considered genomes with mean-fold coverage
151     of at least 20 as high coverage; one historic, seven modern and both outgroup genomes
152     fulfilled this condition (Figure 3a). We identified single nucleotide polymorphisms (SNPs) in
153     each sample independently by comparison with the *P. infestans* reference T30-4 genome
154     (Figure 3b). Thresholds for calling homozygous and heterozygous SNPs were determined
155     from simulated data from high- and low-coverage genomes (Figure 3 – figure supplement 1).
156     We accepted SNPs from low-coverage genomes if the variants had also been called in a high-
157     coverage genome. Inverse cumulative coverage plots indicated how many high- or low-
158     coverage samples were needed to cover different fractions of SNPs (Figure 3c, d). A total of
159     4.5 million non-redundant SNPs were called. Eighty percent of all homozygous SNPs were
160     found in at least eight samples, and only twenty percent of all SNPs were found in fewer than
161     ten strains. Thus, the great majority of polymorphic sites were shared by several strains and
162     thus informative for phylogenetic analyses.

163

164     **A unique type I mtDNA haplotype in nineteenth century *P. infestans* strains**

We reconstructed the mtDNA genomes from historic and modern strains using an iterative mapping assembler (Green et al., 2008) and estimated a phylogenetic tree from complete mtDNA genomes, with one of the *P. mirabilis* mtDNA genomes as outgroup. Previous studies have recognized four *P. infestans* mtDNA haplotype groups (Ia, Ib, IIa and IIb), based on a small number of restriction fragment length polymorphisms (RFLPs) (Carter et al., 1990). Surprisingly, a comparison of the complete mtDNA genomes revealed that the historic samples did not fit into any of these groups, and instead formed an independent clade, called HERB-1 from here on. Among the HERB-1 mtDNA genomes, there were very few differences, with a mean pair-wise difference of only 0.2 bp, compared to 3.9 bp for the modern haplotype I strains, and 9.0 bp for modern haplotype II strains. The origin of HERB-1 relative to haplotypes Ia and Ib could not be unequivocally resolved, and a polytomy was inferred for these three groups or support for branches were low (Figure 4, and Figure 4 – figure supplement 1).

The clonal lineage US-1, with the diagnostic mtDNA haplotype Ib, was the predominant genotype throughout the world until about 1980 (Goodwin et al., 1994). The two US-1 representatives in our material, DDR7602 (Germany) and LBUS5 (South Africa), clustered together with the Ib reference genome and samples P6096 (Peru) and P1362 (Mexico) (Figure 4, and Figure 4 – figure supplement 1), even though these last two samples had not been classified before as US-1 isolates. Although the US-1 genotype is closely related to the herbarium strains, US-1 is not a derivative of HERB-1. Rather, HERB-1 and US-1 are sister groups that share a common ancestor. There are three private substitutions that define the US-1 clade, and two that define the HERB-1 clade. In agreement with the previous report by Ristaino and colleagues (2001), all historic samples had a T at the position diagnostic for haplotype Ib (Figure 4 – figure supplement 2), which distinguishes them from the US-1 lineage, which carries instead a C at this position. In contrast to the previous report (Ristaino et al., 2001), we found no other sequence differences around this diagnostic site.

**Relationship between HERB-1 and modern strains and divergence times**

As the HERB-1 strains were sampled in the nineteenth century, their genomes should harbor fewer substitutions compared to modern strains, which have continued to evolve for over a hundred years. This can be exploited to directly calculate substitution rates and divergence times using the sampling age as tip calibration in a Bayesian framework analysis. Shorter evolutionary time usually translate into branch shortening in phylogenetic trees that include both modern and ancient pathogen strains (Bos et al., 2011).. By calculating the nucleotide distance to the outgroup P17777, all HERB-1 strains were found to show significantly fewer mtDNA substitutions than modern strains with haplotype Ia or Ib (p = 0.0003). Sampling age of the strain and the number of mtDNA substitutions were highly correlated ($r^2 = 0.8$; Figure 5).

Given the correlation between sample age and the number of mtDNA substitutions, a multiple sequence alignment of 12 nearly complete modern and the 13 HERB-1 mtDNA genomes was used as input for a Bayesian framework analysis using algorithms implemented in the software package Beast (Drummond et al., 2012). The molecular clock hypothesis for the modern strains could not be rejected at a 5% significance level (p = 0.12). Therefore, a strict molecular clock and a birth-death tree prior (Stadler, 2010) were used for the Bayesian framework analysis. Collection dates for all herbaria samples and the isolation dates for all modern strains were used as tip calibration points, so that substitution rates per time interval could be calculated (Table 1). Three Markov Chain Monte Carlo (MCMC) runs with 147 million iterations were carried out. Stability of the estimated prior and posterior probability distributions (ESS values >5,000) and likelihood values (ESS values >9,000) were observed in the trace files throughout the merged iterations using the software Tracer (Rambaut and Drummond, 2007). From this procedure, we estimated the mtDNA substitution rate to be 2.4 x $10^{-6}$ per site and year (1.5-3.3 x $10^{-6}$, 95% HPD). This rate resulted in a mean divergence time for *P. infestans* and *P. mirabilis* of 1,318 years ago (ya) (853-1,836 ya 95% HPD) and for *P. infestans* type I and type II mtDNA haplotypes of 460 ya (300-643 ya 95% HPD). The origin of the nineteenth century herbarium clade was estimated to around 182 ya (168-201 ya 95% HPD) (Figure 6, and Table 3).

To understand the evolutionary relationships between historical and modern strains in more detail, we also compared their nuclear genomes. We built phylogenetic trees with high-coverage genomes using maximum parsimony (Figure 7a), maximum likelihood (Figure 7b) and neighbor-joining (Figure 7 – figure supplement 1) methods. We included in the analysis heterozygous biallelic SNPs by random sampling an allele from each of them. In all cases, the HERB-1 representative, M-0182896, formed a distinct, isolated clade that appeared as a robust sister group to US-1 isolates DDR7602 and LBUS5. As a caveat, the random sampling of SNPs at heterozygous positions will inflate bootstrap support. Nevertheless, a heat map (Figure 7c) highlights that the two US-1 strains are both most closely related to HERB-1 and the most distinct among modern strains. Phylogenetic analyses that included the low-coverage genomes (Figure 7 – figure supplement 1) were consistent with a close relationship between the HERB-1 and US-1 lineages.

233

**Ploidy increase in modern strains**

234

The independent diversification of the pandemic HERB-1 and US-1 lineages together with a very recent common ancestor are consistent with both lineages having originated from the same metapopulation. To test whether the global replacement of HERB-1 by US-1 in the twentieth century was due to adaptive mutations, we searched for positively selected genes using PAML (Yang, 2007). We did not find any evidence for genes or sites that had experienced branch-specific positive selection in any of the lineages, only a mosaic pattern with potentially positively selected genes shared across lineages. Alternative scenarios could be that inactivating mutations were more important than non-synonymous substitutions, or that higher overall diversity and re-assortment of beneficial gene variants by recombination contributed to the success of US-1.

Genetic diversity can be increased by polyploidy, which has been reported in isolates of *P. infestans* (e.g., Catal et al., 2010; Daggett et al., 1995), and which has major evolutionary implications for asexual organisms. To estimate ploidy level in our specimens, we investigated the distribution of read counts at biallelic SNPs for high-coverage genomes. In a diploid species, the mean of read counts at heterozygous positions should have a single mode at 0.5, while there should be two modes, 0.33 and 0.67, for triploid genomes, and three modes, 0.25, 0.5 and 0.75 for tetraploid genomes. We compared the observed distributions of read counts with computational simulations of diploid, triploid and tetraploid genomes. Based on the shape and kurtosis of the distributions we concluded that the historic M-0182896 genome was apparently diploid. Of the modern genomes, only NL07434 and P17777 were diploid, whereas the majority, including the two US-1 strains DDR7602 and LBUS5 as well as P13527 and P13626 were triploid. One strain, 06_3928A, even seemed to be tetraploid (Figure 8a, b, and Figure 8 – figure supplement 1). This conclusion was supported by polyploid strains having evidence for triallelic polymorphism at many more sites than M-0182896 (Figure 8c).

To further confirm the ploidy inferences, we classified 40,352 SNPs as derived or ancestral based on information from the outgroup species *P. mirabilis* and *P. ipomoeae.* We then compared the rate of homozygosity at derived alleles in M-0182896 and DDR7602. In agreement with the ploidy difference, M-0182896 had more than twice as many derived homozygous SNPs (8,375) than DDR7602 (3,440), regardless of annotation as synonymous, non-synonymous and non-sense (Figure 9a, b).

266

**Effector genes**

*Phytophthora infestans* secretes a large repertoire of effector proteins, some of which are recognized by plant immune receptors encoded by *R* genes (Haas et al., 2009; Vleeshouwers et al., 2011). These *R* genes occur in wild potato (*Solanum*) species mostly originating from the pathogen center of diversity in Mexico, and have been introduced by breeding into cultivated potato since the beginning of the twentieth century (Hawkes, 1990). The analysis of effector gene sequences in HERB-1 strains should reveal the effector repertoire prior to its disruption by the selective forces imposed by resistance gene breeding. Given that nineteenth century potato cultivars in North America and Europe were fully susceptible to late blight, we presume that they did not yet contain resistance genes that are effective against HERB-1. Conversely, the first *R* genes for *P. infestans* resistance, introduced into cultivated potato only after the dates for our HERB-1 samples, should be effective against HERB-1 strains, which are predicted to carry matching avirulence effector genes. The *R* genes include in particular *R1* to *R4* from *Solanum demissum* (Hawkes, 1990).

To date, ten avirulence effector genes, recognized by ten matching *Solanum R* genes, have been described in *P. infestans* (Vleeshouwers et al., 2011). We first estimated the presence/absence profiles of these effector genes based on the fraction of gene length that was covered by reads ('breadth of coverage') in each high-coverage genome, and by merged reads from low-coverage genomes (Table 4). We deduced the amino acid sequences of these ten effectors using both alignments of reads to the reference genome and *de novo* assemblies. All examined avirulence effector genes except *Avr3b* were present as full-length and intact coding sequences in the historic samples (Table 4), without any frame shift or nonsense mutations. The HERB-1 alleles of *Avr1*, *Avr2*, *Avr3a* and *Avr4* were shared with those of the US-1 strain DDR7602 (Table 5 and source file 1). In conclusion, the *Avr1*, *Avr2*, *Avr3a* and *Avr4* alleles of HERB-1 are intact, presumably functional copies that are identical to ones that can be recognized by the matching *R* genes *R1*, *R2*, *R3a* and *R4* (Armstrong et al., 2005; Gilroy et al., 2011; van Poppel et al., 2008; Vleeshouwers et al., 2011). This is consistent with the expectation that the HERB-1 genotype must have been avirulent on the first potato cultivars that acquired late blight resistance.

We examined in more detail *Avr3a*, which is recognized by *Solanum demissum R3a*. The two major *Avr3a* alleles encode secreted proteins that differ in two amino acids in their effector domains: AVR3a$^{KI}$ and AVR3a$^{EM}$ (Figure 10a, and Figure 10 – figure supplement 1). Only the AVR3a$^{KI}$ type triggers signaling by the resistance protein R3a (Armstrong et al., 2005). The *R3a* gene was introduced into modern potato from *S. demissum* at the beginning of the twentieth century, providing modern potato with resistance against the *P. infestans* strains prevalent at the time (Gebhardt and Valkonen, 2001; Hawkes, 1990; Huang et al., 2005). Strains homozygous for *Avr3a^{EM}*, which avoids *R3a*-mediated detection of the pathogen, appeared later; US-1 isolates lack *Avr3a^{EM}* (Armstrong et al., 2005). Examination of *Avr3a* SNPs in the historic samples only revealed the AVR3a^{KI} allele, whereas both alleles were present in modern samples (Figure 10b). To confirm that the potato hosts of the historic strains lacked the ability to recognize AVR3a^{KI}, we assembled *de novo* short reads from the historic samples and aligned them against the *R3a* sequence from modern potato (Huang et al., 2005). As expected, we only found *R3* homologs that were distinct in sequence from *R3a* (Figure 10c).

The absence of the *Avr3b* effector gene in HERB-1 could be viewed as puzzling, given that the *S. demissum R3* locus was one of the first to be bred into potato. However, *R3b*, the matching *R* gene of *Avr3b*, is within 0.4 cM of *R3a* in the complex *R3* locus (Li et al., 2011). Based on the absence of an *Avr3b* gene in HERB-1, we conclude that initial introgression of the *R3* locus from *S. demissum* was based on the resistance phenotype conferred by the *R3a* gene. The *R3* phenotype scored during the initial introgression must have been the recognition of *Avr3a* by *R3a*, and the presence of *R3b* must have been irrelevant until *P. infestans* strains carrying *Avr3b* emerged.

## Discussion

To characterize the *P. infestans* strain(s) that triggered the Irish potato famine, we have sequenced several mtDNA and nuclear genomes of nineteenth century *P. infestans* strains. Because we wanted to interpret our findings in the context of extant genetic diversity, we investigated several modern strains as well. We could reconstruct the phylogenetic relationship between historic and modern strains using dozens of variants in complete mtDNA genomes, and millions of SNPs in the nuclear genomes, compared to previous work that had to rely on three mtDNA SNPs (May and Ristaino, 2004; Ristaino et al., 2001). The topologies of mtDNA and nuclear phylogenies were very similar, with the nuclear genomes yielding additional insights into differences in heterozygosity, ploidy levels and effector gene complement between historic and modern strains. Contrary to previous inferences (May and Ristaino, 2004; Ristaino et al., 2001), the nineteenth century strains are closely related to the modern US-1 lineage, but are characterized by a single, distinct mtDNA haplotype, HERB-1.

Finally, from estimates of the divergence times of the different lineages, we were able to associate key events in *P. infestans* evolution with historic records of human migration and late blight spread.

**Relationship between historical and modern strains**

Historic strains from different geographic locations all carried a mtDNA haplotype, HERB-1, that had not been recognized before (Figure 4). Although collected over a period of 50 years, the strains were distinguished from each other by few nuclear SNPs, indicating that the nineteenth century outbreak was a true pandemic of a rapidly spreading clonal genotype. That HERB-1 has so far not been found in any modern strain may point to its extinction after the nineteenth century pandemic, possibly associated with the onset of resistance gene breeding in the twentieth century. We cannot, however, exclude that HERB-1 still infects some localized, genetically unimproved host populations, since we have explored only a fraction of current *P. infestans* genetic diversity. With the diagnostic variants we have discovered, one can now probe modern populations to determine whether or not HERB-1 still persists somewhere.

Historic pathogen samples are molecular fossils that can be used as tip calibration points to estimate major divergence events in the evolution of a pathogen (Bos et al., 2011). Using the collection dates of the herbarium samples and isolation dates of the modern *P. infestans* strains, we estimated that type I and type II mtDNA haplotypes diverged close to the beginning of the sixteenth century (Figure 6, and Table 3). This coincides with the first contact between Americans and Europeans in Mexico, which potentially fuelled *P. infestans* population migration and expansion outside its center of diversity. This major event in human history might thus have been responsible for wider dissemination of the *P. infestans* pathogen in the New World, several centuries before its introduction to Europe. In addition, the divergence estimates allowed us to date the split between *P. mirabilis* and *P. infestans* about 1,300 ya. Even though this was firmly during the period of pre-Columbian civilization, what led to their relatively recent speciation remains unknown.

To test the congruence of mtDNA and nuclear phylogenies, we reconstructed phylogenies with over four million nuclear SNPs from high-quality genomes (Figure 7). This confirmed the historic sample M-0182896 as a sister group to US-1 strains, a conclusion that was supported by a broader analysis that included the low-coverage historic samples (Figure 7 – figure supplement 1). The private SNPs shared by the HERB-1 lineage ruled out that US-1 isolates are, as previously proposed (Goodwin et al., 1994), direct descendants of the historic

366 strains. Nevertheless, US-1 is more closely related to the historic strains than to the modern
367 strains that have come to dominate the global population in the past two decades.

368   We therefore propose a revision of the previous scenario, which posited that a single
369 *P. infestans* lineage migrated around 1842 or 1843 from Mexico to North America, from
370 where it was soon transferred to Europe, followed by global dissemination and persistence for
371 over hundred years (Goodwin et al., 1994). Our data make it likely that by the late 1970s,
372 direct descendants of HERB-1 had either become rare or extinct. On the other hand, the close
373 relationship between HERB-1 and US-1 suggests that the US-1 lineage originated from a
374 similar source as HERB-1, with our divergence estimates indicating that the two lineages
375 separated only in the early nineteenth century. Given the much greater genetic diversity at the
376 species' likely origin in Mexico, it seems unlikely that HERB-1 and US-1 spread
377 independently from this region. An alternative scenario is that a small *P. infestans*
378 metapopulation was established at the periphery of its center of origin, or even outside
379 Mexico, possibly in North America, some time before the first global *P. infestans* pandemic.
380 The first lineage to spread from there was HERB-1, which persisted globally for at least half a
381 century. Subsequently, the US-1 lineage expanded and spread, replacing HERB-1 (Figure 11).
382

383 **Genetic and genomic differences between historic and modern strains**
384 Host *R* genes that confer resistance to historic *P. infestans* strains, such as *R3a*, were bred into
385 cultivated potato *Solanum tuberosum* from the wild species *S. demissum* at the beginning of
386 the twentieth century, years after our youngest historic sample had been collected in 1896. In
387 agreement with the products of these *R* genes being able to recognize HERB-1 effectors,
388 HERB-1 strains seem to have only the $Avr3a^{KI}$ allele, which interacts with the product of the
389 *R* gene *R3a* to trigger a host immune response (Armstrong et al., 2005; Huang et al., 2005).
390 Moreover, *de novo* assembly of potato DNA did not provide evidence for the presence of *R3a*
391 in the herbarium hosts, consistent with the narrative of potato breeding (Figure 10c) (Hawkes,
392 1990). While it is uncertain when HERB-1 was displaced by the US-1 lineage, the US-1
393 lineage also carries only the $Avr3a^{KI}$ allele (Armstrong et al., 2005). The origin of the
394 $Avr3a^{EM}$ allele that emerged to high frequency after the breeding of *R3a* into cultivated
395 potatoes remains unclear.

396   A major genomic difference between the HERB-1 and US-1 lineages is the shift in
397 ploidy, from diploid to triploid and even tetraploid (Figure 8, and figure 8 – figure supplement
398 1). Polyploidization could have provided an opportunity for US-1 isolates to enhance allelic
399 diversity in the absence of frequent sexual reproduction, and could thus have contributed to their global success. Asexual reproduction leads to an increase of deleterious mutation in the population due to a lack of meiotic recombination (Felsenstein, 1974). Therefore, if the major selection pressure that led to the replacement of HERB-1 by US-1 was the introduction of resistance gene breeding, greater variation at effector genes in polyploid US-1 strains could have contributed to the replacement of HERB-1 soon after *R* genes from *S. demissum* and other wild species had been introduced into modern potato germplasm.

**Conclusions**

We present the first genome-wide analyses of historic plant pathogen strains. The aDNA in the herbarium samples, which were about 150 years old, was remarkably well conserved, much better than most examples of aDNA from animals and humans, and only comparable with permafrost samples (Miller et al., 2008; Rasmussen et al., 2010).

Our analyses not only highlight how knowledge of the genetics and geographic distribution of modern strains is insufficient to correctly infer the source of historic epidemics (Goodwin et al., 1994), but they also reveal the shortcomings of inferences that are based on a very small number of genetic markers in historic strains (May and Ristaino, 2004; Ristaino et al., 2001). With our much richer dataset, we could demonstrate that the late blight outbreaks during the nineteenth century were a pandemic caused by a single *P. infestans* lineage, but that this lineage was not the direct ancestor of the one that had come to dominate the global *P. infestans* population during much of the twentieth century. Infected plant specimens stored in herbaria around the world are thus a largely untapped source to learn about events that affected millions of people during our recent history.

**Material and methods**

**Herbarium sampling**

Plant specimens were sent to the Senckenberg Museum in Frankfurt am Main by the Botanische Staatssammlung München and the Kew Royal Botanical Gardens, where potato and tomato leaves with lesions indicative of *P. infestans* infection were sampled, retrieving both the lesions and healthy surrounding tissue. Sampling was carried out under sterile conditions in a laboratory with no prior exposure to *P. infestans*. Samples were subsequently sent to the Palaeogenetics laboratory at the University of Tübingen.

**DNA extraction and sequencing library preparation**

Preamplification steps of historic samples were performed in clean room facilities with no prior exposure to *P. infestans* DNA. Samples were extracted following the protocol of (Kistler, 2012), using 380 to 500 µg of each sample. Tissue was crushed with mortal and pestle, 1.2 ml extraction buffer (1% SDS, 10 mM Tris pH 8.0, 5 mM NaCl, 50 mM DTT, 0.4 mg/mL proteinase K, 10 mM EDTA, 2.5 mM *N*-phenacylthiazolium bromide) was added, and samples were incubated over night at 37°C with constant agitation. A modified protocol with the Qiagen Plant DNEasy Mini kit was then used to purify the extracted DNA.

Two independent Illumina sequencing libraries were created for each DNA extract. In the first library, C-to-T damage caused by deamination of cytosines (Hofreiter et al., 2001) was not repaired. Twenty µl of each DNA extract, extraction blank control and water library blank control were converted into sequencing libraries as described (Meyer and Kircher, 2010) with modifications for aDNA (Meyer et al., 2012). To avoid potential sequencing artifacts caused by DNA damage, a second library was made from, 30 µl of each DNA extract, extraction blank control and water library blank control, and treated with uracil-DNA glycosylase (UDG) and Endonuclease VIII before conversion into sequencing libraries (Briggs et al., 2010). Each library received sample-specific double indices after preparation via amplification with two 'index' PCR primers (Meyer et al., 2012). Indexed libraries were individually amplified in 100 µl reactions containing 5 µl library template, 2 units of AccuPrime Pfx DNA polymerase (Invitrogen), 1 unit of 10×PCR Mix and 0.3 µM primers spanning the index sequences of the libraries. The following thermal profile was used: 2-min initial denaturation at 95°C, 2 or 3 cycles consisting of 15 sec denaturation at 95°C, a 30-sec annealing at 60°C and a 2-min elongation at 68°C, and a 5-min final elongation at 68°C. Amplified products were purified and quantified on an Agilent 2100 Bioanalyzer DNA 1000 chip.

DNA extracts of the modern *P. infestans* samples P17721, P10650, P6096, P11633, P10127, P9464, P12204 and P13626 and *P. mirabilis* P7722 were obtained from the World Phytophthora and Oomycete Genetic Resource Collection, fragmented to 300 bp using a S220 Covaris instrument according to the manufacturers' protocol (Duty cycle 10%, intensity 4, cycles per burst 200, time (sec) 120), and converted into sequencing libraries following the above steps as described for the historic samples without UDG treatment (Kircher, 2012; Meyer et al., 2012). For *P. mirabilis* PIC99114 and *P. ipomoeae* PIC99167, genomic DNA used before (Cooke et al., 2012; Raffaele et al., 2010b) was converted into Illumina

464 sequencing libraries. Genomic DNA from the other modern strains was isolated as described
465 (Cooke et al., 2012).

466         Libraries were sequenced on Illumina GAIIx, HiSeq 2000 or MiSeq instruments,
467 (Table 2). To estimate the fraction of *P. infestans* aDNA in the herbarium samples, we
468 performed pilot sequencing. Once the samples with the highest amount of *P. infestans* were
469 identified, production runs were carried out on an Illumina HiSeq 2000 instrument. For *P.*
470 *infestans* 06_3928A analysis, we used publicly available short reads (ENA ERP002420).

471

472 **Read mapping and SNP calling**

473 Reads for the herbarium samples were de-indexed as described (Kircher, 2012). Forward and
474 reverse reads were merged after adapter trimming, requiring at least 11 nucleotides overlap
475 (Burbano et al., 2010). Only fragments that that allowed merging of reads were used in
476 subsequent analyses. Merged reads were mapped to the *P. infestans* T30-4 reference genome
477 (Haas et al., 2009) with BWA, allowing two gaps and without seeding (Li and Durbin, 2009).
478 PCR duplicates were identified based on read start and end alignment coordinates. For each
479 cluster of duplicates a consensus sequence was calculated as described (Kircher, 2012). Only
480 reads with a Phred-like mapping quality score of at least 30 were used further. Alignments
481 were converted to BAM files (Li et al., 2009). For modern strains, single reads were mapped
482 in a similar manner, and subsequent processing was performed as described (Cooke et al.,
483 2012).

484         SNPs in the herbarium samples were called by independently comparing each strain
485 with the *P. infestans* T30-4 genome. Raw allele counts for each position were obtained using
486 pileup from SAMtools (Li et al., 2009). To decide the cutoffs for SNP calling we resorted to
487 simulations. Reads from two 50-fold and 3-fold coverage genomes were simulated using the
488 pIRS software (Hu et al., 2012), with empirical base-calling and GC%-depth profiles trained
489 on five modern *P. infestans* genomes (P13527, P13626, 06_3928A, NL07434 and P17777).
490 Based on the simulated data we called both homo and heterozygous SNPs employing
491 different cutoffs for SNP concordance (Figure 3 – figure supplement 1). Genotypes calls were
492 classified as high quality if coverage was at least 10. We also considered low-quality SNPs, if
493 a high-quality SNP call had been made in a different strain, using specific coverage cutoffs
494 for rescuing low-quality SNPs (Figure 3 – figure supplement 1). We calculated sensitivity
495 and accuracy of SNP calls for different combination of cutoffs and selected the following
496 criteria:

497        •   Minimum coverage of 10 for high quality calls.

498       •   Concordance >= 80% for homozygous SNPs.

499       •   Concordance between 20-80% for heterozygous SNPs.

500       •   Minimum coverage of 3 to rescue low-quality SNPs.

501       We called synonymous, nonsynonymous and nonsense polymorphisms employing

502 snpEff (Cingolani et al., 2012).

503

504 **Mitochondrial DNA (mtDNA) assembly and phylogenetic analyses**

505 Fragments that could be aligned to any of the four reference haplotypes (Ia, IIa, Ib and IIb)

506 were used to assemble mtDNA genomes. For each strain four different assemblies (one for

507 each reference haplotype) were built using an iterative mapping assembly program (Burbano

508 et al., 2010; Green et al., 2008). Only positions with coverage of at least 3 were included in

509 the assemblies. The four assemblies were aligned using Kalign (Lassmann and Sonnhammer,

510 2005) with default parameters, and a consensus assembly was generated based on the

511 alignment.

512       The 1.8 kb insertion present in haplotype II was not considered for phylogenetic

513 reconstruction. The mtDNA phylogeny was built using a maximum parsimony and a

514 maximum likelihood tree using MEGA5 (Tamura et al., 2011). For both, positions with less

515 than 90% site coverage were eliminated. There were a total of 24,560 positions in the final

516 dataset, compared to the multiple sequence alignment length (37,762 bp). For the maximum

517 likelihood reconstruction we used the Hasegawa-Kishino-Yano (HKY) model assuming that a

518 certain fraction of sites are evolutionarily invariable. The model was selected using MEGA5

519 (Tamura et al., 2011).

520

521 **Nuclear genome phylogenetic analyses**

522 We reconstructed the nuclear phylogeny for the high-coverage samples alone and for all

523 samples together independently, using maximum parsimony and maximum likelihood

524 approaches. We built the neighbor-joining trees based on a genetic distance matrix calculated

525 from both homo- and heterozygous SNPs (Xu et al., 2012). For the high-coverage genomes

526 we used only the SNPs positions with complete information in all samples. For the low-

527 coverage genomes we used homo- and heterozygous SNPs, and allowing for missing data. So

528 that we could include heterozygous SNPs in the analysis, we randomly sampled one allele at

529 each site. The maximum parsimony trees were calculated with 100 replicates in MEGA5

530 using the Close-Neighbor-Interchange algorithm with search level 0, in which the initial trees

531 were obtained with the random addition of sequences (10 replicates). All positions with less
532 than 95% site coverage were eliminated (Tamura et al., 2011). For the high-coverage
533 genomes-only analysis, all positions with less than 85% site coverage were eliminated. For
534 the all-sample analysis the threshold was lowered to 80%. Maximum likelihood trees were
535 built using RaxML 7.0.4 with 100 replicates using the rapid bootstrap algorithm (Stamatakis,
536 2006).
537
538 **Effector analyses**
539 To address presence and absence polymorphisms of effectors, we used a previously published
540 pipeline (Raffaele et al., 2010a). We calculated the fraction of each gene that was covered by
541 reads ('breadth of coverage') for each strain. We regarded zero breadth of coverage as
542 absence of the gene. For herbarium and modern samples with genome-wide coverage depth
543 less than 20x, we merged BAM files from each strain into a single BAM file, and used this to
544 estimated breadth of coverage.
545      We also tested for presence/absence polymorphisms of RXLR effector genes between
546 herbarium samples and modern strains using *de novo* assembly of short reads. First, we
547 extracted 140 bp fragments from M-0182896 merged reads, and assembled these with
548 SOAPdenovo v1.05 (Luo et al., 2012). We aligned the six-frame translation of resulting
549 contigs to a non-redundant protein database using blastx (Altschul et al., 1990). Most contigs
550 matched proteins from either potato, *Solanum tuberosum*, or to microbial species *P. infestans*,
551 *Pantoea vagans* and *Pseudomonas* sp. To focus on *P. infestans*, we aligned fragments that
552 were at least 140 bp to the genomes of *P. infestans*, *S. tuberosum*, *P. vagans*, *P. syringae* pv.
553 *syringae*, and *P. fluorescens* with blastn. We extracted fragments that aligned the *P. infestans*
554 genomic regions encoding RXLR effector genes, but over at most 90 bp. These and
555 unmapped fragments were *de novo* assembled with SOAPdenovo v1.05. A *k*-mer size of 67
556 was deemed optimal, because it resulted in the highest coverage of *Avr1*, *Avr2* and *Avr3a*, and
557 resulted in the largest number of RXLR proteins with TBLASTN hits (Figure 10 – figure
558 supplement 1). We obtained partial sequences of *Avr4* and *Avrblb1*. We visually evaluated
559 BWA alignments of M-0182896 in the *Avr4* and *Avrblb1* genomic regions and identified T30-
560 4 sequences uncovered by alignments using Integrative Genomics Viewer (Robinson et al.,
561 2011). We then identified T30-4 genomic regions with at least 99% similarity to these
562 uncovered regions. In BWA, if reads match several genomic regions, one genomic location is
563 randomly chosen as default (Li and Durbin, 2009). Thus, it is possible that BWA alignment
564 distributes reads coming from the same gene across several, closely related genes in the target

565 genome. We assembled such reads that mapped to closely related sequences in the reference
566 genome together with the partial sequences of *Avr4* and *Avrblb1* using Geneious® Pro 5.6.3
567 to obtain full-length sequences of these *Avr* genes.
568
569 **Selection tests**
570 Homozygous SNPs from modern *P. infestans* strains EC3527, EC3626, NL07434, 06_3928A,
571 DDR7602, LBUS5, P17777 and the historic strain M-0182896 were used for selection tests.
572 Gene sequences were converted into amino acid sequences using EMBOSS tools (Rice et al.,
573 2000), and Pla2Nal v14 (Suyama et al., 2006) was used to convert protein alignments to
574 codon alignments. The codeml module of PAML package v4.6 (Yang, 2007) was used for
575 positive selection studies with site models M7 (parameters NSsites = 7, fix_omega = 0,
576 omega = 2 and kappa = 3) and M8 (NSsites = 8, fix_omega = 0, omega = 2 and kappa = 3). A
577 5% level of significance was established with Likelihood ratio test. Genes were considered to
578 be under positive selection if at least one site was found to be under selection with a Bayes
579 Empirical Bayes confidence >95%.
580
581 **Ploidy analyses**
582 To estimate ploidy levels, we assessed the distributions of read counts at biallelic SNPs. For
583 diploid species, the mean frequency of reads for each allele at non-homozygous sites is 1/2,
584 while we expect two modes for triploid genomes, at 1/3 and 2/3, and four modes for tetraploid
585 genomes, at 1/4, 1/2 and 3/4 (Figure 8a). We simulated genomes with different ploidy levels
586 using pIRS (Hu et al., 2012), based on two strains, *P. infestans* T30-4 and EC3527. The SNPs
587 used for the construction of two simulated chromosomes were determined with SAMtools
588 v0.1.8 mpileup and bcftools v0.1.17 (Li et al., 2009). For the diploid genome, we simulated
589 10x coverage reads for each of two different chromosomes. For the triploid genome, we
590 merged simulated 5x and 15x coverage reads from two different chromosomes. For the
591 tetraploid genome, we merged simulated 10x coverage reads from two different chromosomes
592 (Figure 8b). Next, we aligned the simulated reads to the *P. infestans* T30-4 reference genome
593 with BWA and called heterozygous SNPs under the following criteria: minimum coverage of
594 10 for high-quality calls, and concordance between 20-80% for heterozygous SNPs. Since
595 tetraploid species are considered to be a mixtures of two ratio, we mixed SNPs from the 20x
596 coverage diploid reads and the 20x coverage tetraploid reads in following ratios: 0:100, 10:90,
597 20:80, 30:70, 40:60, 50:50, 60:40, 70:30, 80:20, 90:10 and 100:0. Finally, we estimated
598 frequency of reads assigning each allele at each SNP position. Based on shapes, standard

599  deviation, skewedness and kurtosis of the observed distributions and comparison with the
600  simulated distributions, we classified the tested *P. infestans* genomes as diploid, triploid and
601  tetraploid.
602
603  **Substitution rates and divergence times for *P. infestans***
604  In order to test whether we can detect a temporal signal in the ancient *P. infestans* mtDNA
605  sequences compared to modern strains, i.e. shorter branches in the ancient strains compared to
606  the modern ones, we calculated the nucleotide distance as the number of substitutions
607  between HERB-1, haplotype Ia and haplotype Ib mtDNA genomes to the outgroup P17777.
608  The analysis involved 19 nucleotide sequences. All positions with less than 90% site coverage
609  were eliminated, resulting in 34,174 informative positions. The samples were subsequently
610  grouped into ancient and modern strains. The ancient and modern nucleotide distances were
611  significantly different (Mann-Whitney U-test, p = 0.0003). We furthermore correlated the
612  nucleotide distance of HERB-1, haplotype Ia and haplotype Ib mtDNA genomes to the
613  outgroup P17777 with the tip age of each sample.
614      To estimate divergence times of *P. infestans* strains, substitution rates were calculated
615  in a Bayesian framework analysis using the software package BEAST 1.7.5 (Drummond et
616  al., 2012). A multiple sequence alignment that included all 12 nearly complete modern *P.
617  infestans* mtDNA sequences plus all 13 herbaria samples was used as input. In order to test if
618  the mtDNAs evolved clock like a likelihood ratio test was performed in MEGA5 (Tamura et
619  al., 2011) by comparing the maximum likelihood (ML) value for the given topology using
620  only the modern strains with and without the molecular clock constraints. The null hypothesis
621  of equal evolutionary rate throughout the tree was not rejected at a 5% significance level (P =
622  0.115). All positions containing gaps and missing data were eliminated resulting in a total of
623  22,591 positions in the final dataset.
624      As a result a strict molecular clock and the HKY sequence evolution model were used
625  for the Bayesian framework. For the tree prior, five different models were tested including
626  four coalescence models: constant size, expansion growth, exponential growth, logistic
627  growth and a epidemiology birth-death model (Stadler, 2010). For each tree prior, three
628  MCMC runs were carried out with 10,000,000 iterations each and subsequently merged using
629  LogCombiner 1.7.5 from the BEAST package. Resulting ESS values and overall posterior
630  likelihoods were compared using the software Tracer (Rambaut and Drummond, 2007). The
631  birth-death model gave the highest ESS values and posterior likelihood and was therefore
632  chosen for the subsequent dating analysis. The collection dates for all herbaria samples as 633 well as the isolation dates for all modern strains were used as tip calibration points (Table 1).
634 Three MCMC runs were carried out with 50,000,000 iterations each, sampling every 10,000
635 steps. The first 1,000,000 iterations were discarded as burn-in resulting in a total of
636 147,000,000 iterations.

## Bibliography

638 Altschul, S.F., Gish, W., Miller, W., Myers, E.W., and Lipman, D.J. (1990). Basic local
639     alignment search tool. J. Mol. Biol. *215*, 403-410.
640 Armstrong, M.R., Whisson, S.C., Pritchard, L., Bos, J.I., Venter, E., Avrova, A.O., Rehmany,
641     A.P., Bohme, U., Brooks, K., Cherevach, I., et al. (2005). An ancestral oomycete locus
642     contains late blight avirulence gene *Avr3a*, encoding a protein that is recognized in the host
643     cytoplasm. Proc. Natl. Acad. Sci. USA *102*, 7766-7771.
644 Bos, K.I., Schuenemann, V.J., Golding, G.B., Burbano, H.A., Waglechner, N., Coombes,
645     B.K., McPhee, J.B., DeWitte, S.N., Meyer, M., Schmedes, S., et al. (2011). A draft
646     genome of *Yersinia pestis* from victims of the Black Death. Nature *478*, 506-510.
647 Bos, K.I., Stevens, P., Nieselt, K., Poinar, H.N., Dewitte, S.N., and Krause, J. (2012). *Yersinia*
648     *pestis*: new evidence for an old infection. PLoS ONE *7*, e49803.
649 Bourke, P.M.A. (1964). Emergence of potato blight, 1843-46. Nature *203*, 805-808.
650 Briggs, A.W., Stenzel, U., Johnson, P.L., Green, R.E., Kelso, J., Prüfer, K., Meyer, M.,
651     Krause, J., Ronan, M.T., Lachmann, M., et al. (2007). Patterns of damage in genomic
652     DNA sequences from a Neandertal. Proc. Natl. Acad. Sci. USA *104*, 14616-14621.
653 Briggs, A.W., Stenzel, U., Meyer, M., Krause, J., Kircher, M., and Pääbo, S. (2010). Removal
654     of deaminated cytosines and detection of in vivo methylation in ancient DNA. Nucleic
655     Acids Res. *38*, e87.
656 Brotherton, P., Endicott, P., Sanchez, J.J., Beaumont, M., Barnett, R., Austin, J., and Cooper,
657     A. (2007). Novel high-resolution characterization of ancient DNA reveals C > U-type base
658     modification events as the sole cause of post mortem miscoding lesions. Nucleic Acids
659     Res. *35*, 5717-5728.
660 Burbano, H.A., Hodges, E., Green, R.E., Briggs, A.W., Krause, J., Meyer, M., Good, J.M.,
661     Maricic, T., Johnson, P.L., Xuan, Z., et al. (2010). Targeted investigation of the Neandertal
662     genome by array-based sequence capture. Science *328*, 723-725.

663    Carter, D.A., Archer, S.A., Buck, K.W., Shaw, D.S., and Shattock, R.C. (1990). Restriction
664        fragment length polymorphisms of mitochondrial DNA of *Phytophthora infestans*. Mycol.
665        Res. *94*, 1123-1128.

666    Catal, M., King, L., Tumbalam, P., Wiriyajitsomboon, P., Kirk, W.W., and Adams, G.C.
667        (2010). Heterokaryotic nuclear conditions and a heterogeneous nuclear population are
668        observed by flow cytometry in *Phytophthora infestans*. Cytometry. Part A : the journal of
669        the International Society for Analytical Cytology *77*, 769-775.

670    Cingolani, P., Platts, A., Wang le, L., Coon, M., Nguyen, T., Wang, L., Land, S.J., Lu, X.,
671        and Ruden, D.M. (2012). A program for annotating and predicting the effects of single
672        nucleotide polymorphisms, SnpEff: SNPs in the genome of Drosophila melanogaster strain
673        *w^{1118}; iso-2; iso-3*. Fly *6*, 80-92.

674    Cooke, D.E., Cano, L.M., Raffaele, S., Bain, R.A., Cooke, L.R., Etherington, G.J., Deahl,
675        K.L., Farrer, R.A., Gilroy, E.M., Goss, E.M., et al. (2012). Genome analyses of an
676        aggressive and invasive lineage of the Irish potato famine pathogen. PLoS Pathog. *8*,
677        e1002940.

678    Daggett, S.S., Knighton, J.E., and Therrien, C.D. (1995). Polyploidy among isolates of
679        *Phytophthora infestans* from Eastern Germany. J. Phytopathol. *143*, 419-422.

680    de Bary, H.A. (1876). Researches into the nature of the potato fungus, *Phytophthora
681        infestans*. J. Roy. Agric. Soc. Engl. Ser. 2 *12*, 239-269.

682    Drummond, A.J., Suchard, M.A., Xie, D., and Rambaut, A. (2012). Bayesian phylogenetics
683        with BEAUti and the BEAST 1.7. Mol. Biol. Evol. *29*, 1969-1973.

684    Felsenstein, J. (1974). The evolutionary advantage of recombination. Genetics *78*, 737-756.

685    Fisher, M.C., Henk, D.A., Briggs, C.J., Brownstein, J.S., Madoff, L.C., McCraw, S.L., and
686        Gurr, S.J. (2012). Emerging fungal threats to animal, plant and ecosystem health. Nature
687        *484*, 186-194.

688    Flier, W.G., Grunwald, N.J., Kroon, L.P., Sturbaum, A.K., van den Bosch, T.B., Garay-
689        Serrano, E., Lozoya-Saldana, H., Fry, W.E., and Turkensteen, L.J. (2003). The population
690        structure of *Phytophthora infestans* from the Toluca Valley of Central Mexico suggests
691        genetic differentiation between populations from cultivated potato and wild *Solanum*
692        species. Phytopathology *93*, 382-390.

693    Fry, W. (2008). *Phytophthora infestans*: the plant (and *R* gene) destroyer. Mol. Plant Pathol.
694        *9*, 385-402.

695    Fry, W., Grünwald, N., D, C., A, M., G, F., and K, C. (2009). Population genetics and
696        population diversity of *Phytophthora infestans*. In Oomycete Genetics and Genomics:

697     Diversity, Interactions, and Research Tools, K. Lamour, and S. Kamoun, eds. (Hoboken,
698     NJ, Wiley-Blackwell), pp. 139-164.

699   Fry, W.E., Goodwin, S.B., Matuszak, J.M., Spielman, L.J., Milgroom, M.G., and Drenth, A.
700     (1992). Population genetics and intercontinental migrations of *Phytophthora infestans*.
701     Annu. Rev. Phytopathol. *30*, 107-129.

702   Gavino, P.D., and Fry, W.E. (2002). Diversity in and evidence for selection on the
703     mitochondrial genome of *Phytophthora infestans*. Mycologia *94*, 781-793.

704   Gebhardt, C., and Valkonen, J.P. (2001). Organization of genes controlling disease resistance
705     in the potato genome. Annu. Rev. Phytopathol. *39*, 79-102.

706   Gilroy, E.M., Breen, S., Whisson, S.C., Squires, J., Hein, I., Kaczmarek, M., Turnbull, D.,
707     Boevink, P.C., Lokossou, A., Cano, L.M., et al. (2011). Presence/absence, differential
708     expression and sequence polymorphisms between *PiAVR2* and *PiAVR2-like* in
709     *Phytophthora infestans* determine virulence on *R2* plants. New Phytol. *191*, 763-776.

710   Goodwin, S.B., Cohen, B.A., and Fry, W.E. (1994). Panglobal distribution of a single clonal
711     lineage of the Irish potato famine fungus. Proc. Natl. Acad. Sci. USA *91*, 11591-11595.

712   Green, R.E., Malaspinas, A.S., Krause, J., Briggs, A.W., Johnson, P.L., Uhler, C., Meyer, M.,
713     Good, J.M., Maricic, T., Stenzel, U., et al. (2008). A complete Neandertal mitochondrial
714     genome sequence determined by high-throughput sequencing. Cell *134*, 416-426.

715   Grünwald, N.J., and Flier, W.G. (2005). The biology of *Phytophthora infestans* at its center of
716     origin. Annu. Rev. Phytopathol. *43*, 171-190.

717   Haas, B.J., Kamoun, S., Zody, M.C., Jiang, R.H., Handsaker, R.E., Cano, L.M., Grabherr, M.,
718     Kodira, C.D., Raffaele, S., Torto-Alalibo, T., et al. (2009). Genome sequence and analysis
719     of the Irish potato famine pathogen *Phytophthora infestans*. Nature *461*, 393-398.

720   Hawkes, J.G. (1990). The Potato: Evolution, Biodiversity and Genetic Resources
721     (Washington, D.C., Smithsonian Institution Press).

722   Hofreiter, M., Jaenicke, V., Serre, D., Haeseler Av, A., and Pääbo, S. (2001). DNA sequences
723     from multiple amplifications reveal artifacts induced by cytosine deamination in ancient
724     DNA. Nucleic Acids Res. *29*, 4793-4799.

725   Hu, X., Yuan, J., Shi, Y., Lu, J., Liu, B., Li, Z., Chen, Y., Mu, D., Zhang, H., Li, N., et al.
726     (2012). pIRS: Profile-based Illumina pair-end reads simulator. Bioinformatics *28*, 1533-
727     1535.

728   Huang, S., van der Vossen, E.A., Kuang, H., Vleeshouwers, V.G., Zhang, N., Borm, T.J., van
729     Eck, H.J., Baker, B., Jacobsen, E., and Visser, R.G. (2005). Comparative genomics
730     enabled the isolation of the *R3a* late blight resistance gene in potato. Plant J. *42*, 251-261.

731    Kamoun, S., Hraber, P., Sobral, B., Nuss, D., and Govers, F. (1999). Initial assessment of
732        gene diversity for the oomycete pathogen *Phytophthora infestans* based on expressed
733        sequences. Fungal Genet. Biol. *28*, 94-106.

734    Kircher, M. (2012). Analysis of high-throughput ancient DNA sequencing data. Methods
735        Mol. Biol. *840*, 197-228.

736    Kistler, L. (2012). Ancient DNA extraction from plants. Methods Mol. Biol. *840*, 71-79.

737    Kroon, L.P., Bakker, F.T., van den Bosch, G.B., Bonants, P.J., and Flier, W.G. (2004).
738        Phylogenetic analysis of *Phytophthora* species based on mitochondrial and nuclear DNA
739        sequences. Fungal Genet. Biol. *41*, 766-782.

740    Lassmann, T., and Sonnhammer, E.L. (2005). Kalign--an accurate and fast multiple sequence
741        alignment algorithm. BMC Bioinformatics *6*, 298.

742    Li, G., Huang, S., Guo, X., Li, Y., Yang, Y., Guo, Z., Kuang, H., Rietman, H., Bergervoet,
743        M., Vleeshouwers, V.G., et al. (2011). Cloning and characterization of *R3b*; members of
744        the *R3* superfamily of late blight resistance genes show sequence and functional
745        divergence. Mol. Plant Microbe Interact. *24*, 1132-1142.

746    Li, H., and Durbin, R. (2009). Fast and accurate short read alignment with Burrows-Wheeler
747        transform. Bioinformatics *25*, 1754-1760.

748    Li, H., Handsaker, B., Wysoker, A., Fennell, T., Ruan, J., Homer, N., Marth, G., Abecasis, G.,
749        and Durbin, R. (2009). The Sequence Alignment/Map format and SAMtools.
750        Bioinformatics *25*, 2078-2079.

751    Luo, R., Liu, B., Xie, Y., Li, Z., Huang, W., Yuan, J., He, G., Chen, Y., Pan, Q., Liu, Y., et al.
752        (2012). SOAPdenovo2: an empirically improved memory-efficient short-read de novo
753        assembler. GigaScience *1*, 18.

754    May, K.J., and Ristaino, J.B. (2004). Identity of the mtDNA haplotype(s) of *Phytophthora
755        infestans* in historical specimens from the Irish potato famine. Mycol. Res. *108*, 471-479.

756    Meyer, M., and Kircher, M. (2010). Illumina sequencing library preparation for highly
757        multiplexed target capture and sequencing. Cold Spring Harb. Protoc. *2010*, pdb prot5448.

758    Meyer, M., Kircher, M., Gansauge, M.T., Li, H., Racimo, F., Mallick, S., Schraiber, J.G., Jay,
759        F., Prufer, K., de Filippo, C., et al. (2012). A high-coverage genome sequence from an
760        archaic Denisovan individual. Science *338*, 222-226.

761    Miller, W., Drautz, D.I., Ratan, A., Pusey, B., Qi, J., Lesk, A.M., Tomsho, L.P., Packard,
762        M.D., Zhao, F., Sher, A., et al. (2008). Sequencing the nuclear genome of the extinct
763        woolly mammoth. Nature *456*, 387-390.

764    Pääbo, S. (1989). Ancient DNA: extraction, characterization, molecular cloning, and
765        enzymatic amplification. Proc. Natl. Acad. Sci. USA *86*, 1939-1943.

766    Pääbo, S., Poinar, H., Serre, D., Jaenicke-Despres, V., Hebler, J., Rohland, N., Kuch, M.,
767        Krause, J., Vigilant, L., and Hofreiter, M. (2004). Genetic analyses from ancient DNA.
768        Annu. Rev. Genet. *38*, 645-679.

769    Perez, W.G., Gamboa, J.S., Falcon, Y.V., Coca, M., Raymundo, R.M., and Nelson, R.J.
770        (2001). Genetic structure of peruvian populations of *Phytophthora infestans*.
771        Phytopathology *91*, 956-965.

772    Potato Genome Sequencing Consortium (2011). Genome sequence and analysis of the tuber
773        crop potato. Nature *475*, 189-195.

774    Raffaele, S., Farrer, R.A., Cano, L.M., Studholme, D.J., MacLean, D., Thines, M., Jiang,
775        R.H., Zody, M.C., Kunjeti, S.G., Donofrio, N.M., et al. (2010a). Genome evolution
776        following host jumps in the Irish potato famine pathogen lineage. Science *330*, 1540-1543.

777    Raffaele, S., Win, J., Cano, L.M., and Kamoun, S. (2010b). Analyses of genome architecture
778        and gene expression reveal novel candidate virulence factors in the secretome of
779        *Phytophthora infestans*. BMC Genomics *11*, 637.

780    Rambaut, A., and Drummond, A. (2007). Tracer v1.4. http://beast.bio.ed.ac.uk/Tracer

781    Rasmussen, M., Li, Y., Lindgreen, S., Pedersen, J.S., Albrechtsen, A., Moltke, I., Metspalu,
782        M., Metspalu, E., Kivisild, T., Gupta, R., et al. (2010). Ancient human genome sequence of
783        an extinct Palaeo-Eskimo. Nature *463*, 757-762.

784    Reader, J. (2009). Potato: A History of the Propitious Esculent (New Haven, Yale University
785        Press).

786    Rice, P., Longden, I., and Bleasby, A. (2000). EMBOSS: the European Molecular Biology
787        Open Software Suite. Trends Genet. *16*, 276-277.

788    Ristaino, J.B., Groves, C.T., and Parra, G.R. (2001). PCR amplification of the Irish potato
789        famine pathogen from historic specimens. Nature *411*, 695-697.

790    Robinson, J.T., Thorvaldsdottir, H., Winckler, W., Guttman, M., Lander, E.S., Getz, G., and
791        Mesirov, J.P. (2011). Integrative genomics viewer. Nat. Biotechnol. *29*, 24-26.

792    Stadler, T. (2010). Sampling-through-time in birth-death trees. J Theor Biol *267*, 396-404.

793    Stamatakis, A. (2006). RAxML-VI-HPC: maximum likelihood-based phylogenetic analyses
794        with thousands of taxa and mixed models. Bioinformatics *22*, 2688-2690.

795    Suyama, M., Torrents, D., and Bork, P. (2006). PAL2NAL: robust conversion of protein
796        sequence alignments into the corresponding codon alignments. Nucleic Acids Res. *34*,
797        W609-612.

798    Tamura, K., Peterson, D., Peterson, N., Stecher, G., Nei, M., and Kumar, S. (2011). MEGA5:
799        molecular evolutionary genetics analysis using maximum likelihood, evolutionary
800        distance, and maximum parsimony methods. Mol. Biol. Evol. *28*, 2731-2739.

801    The Tomato Genome Consortium (2012). The tomato genome sequence provides insights into
802        fleshy fruit evolution. Nature *485*, 635-641.

803    Tooley, P.W., Fry, W.E., and Gonzalez, M.J.V. (1985). Isozyme characterization of sexual
804        and asexual *Phytophthora infestans* populations. J. Hered. *76*, 431-435.

805    Turner, R.S. (2005). After the famine: Plant pathology, *Phytophthora infestans*, and the late
806        blight of potatoes, 1845-1960. Hist. Stud. Phys. Biol. Sci. *35*, 341-370.

807    van Poppel, P.M., Guo, J., van de Vondervoort, P.J., Jung, M.W., Birch, P.R., Whisson, S.C.,
808        and Govers, F. (2008). The *Phytophthora infestans* avirulence gene *Avr4* encodes an
809        RXLR-dEER effector. Mol. Plant Microbe Interact. *21*, 1460-1470.

810    Vleeshouwers, V.G., Raffaele, S., Vossen, J.H., Champouret, N., Oliva, R., Segretin, M.E.,
811        Rietman, H., Cano, L.M., Lokossou, A., Kessel, G., et al. (2011). Understanding and
812        exploiting late blight resistance in the age of effectors. Annu. Rev. Phytopathol. *49*, 507-
813        531.

814    Xu, X., Liu, X., Ge, S., Jensen, J.D., Hu, F., Li, X., Dong, Y., Gutenkunst, R.N., Fang, L.,
815        Huang, L., et al. (2012). Resequencing 50 accessions of cultivated and wild rice yields
816        markers for identifying agronomically important genes. Nat. Biotechnol. *30*, 105-111.

817    Yang, Z. (2007). PAML 4: phylogenetic analysis by maximum likelihood. Mol. Biol. Evol.
818        *24*, 1586-1591.

819

## Acknowledgments

821 We are indebted to Bryn Dentinger, curator of the herbarium of the Kew Royal Botanical
822 Gardens, and to Dagmar Triebel, curator of the herbarium of the Botanische Staatssammlung
823 München, for providing the historic specimens used in this study. We are grateful to Mike
824 Coffey for discussion of strain selection for genome analyses and providing genomic DNA
825 preparations, Tahir Ali for help with phylogenetic tests, Jodie Pike for sequencing support,
826 and Mike Coffey, David Cooke, Geert Kessel, Adele McLeod, Ricardo Oliva and Vivianne
827 Vlesshouwers for *Phytophthora* strains. We thank Axel Künstner, Dan Koenig, Jorge

830

831

832

833

# Tables

**Table 1.** Provenance of *P. infestans* samples.

| | ID | Country of origin | Collection year | Host species | Reference* |
|---|---|---|---|---|---|
| **Herbarium samples** | KM177500 | England | 1845 | *Solanum tuberosum* | 1 |
| | KM177513 | Ireland | 1846 | *Solanum tuberosum* | 1 |
| | KM177502 | England | 1846 | *Solanum tuberosum* | 1 |
| | KM177497 | England | 1846 | *Solanum tuberosum* | 1 |
| | KM177514 | Ireland | 1847 | *Solanum tuberosum* | 1 |
| | KM177548 | England | 1847 | *Solanum tuberosum* | 1 |
| | KM177507 | England | 1856 | *Petunia hybrida* | 1 |
| | M-0182898 | Germany | Before 1863 | *Solanum tuberosum* | 2 |
| | KM177509 | England | 1865 | *Solanum tuberosum* | 1 |
| | M-0182900 | Germany | 1873 | *Solanum lycopersicum* | 2 |
| | M-0182907 | Germany | 1875 | *Solanum tuberosum* | 1 |
| | KM177517 | Wales | 1875 | *Solanum tuberosum* | 1 |
| | M-0182897 | USA | 1876 | *Solanum lycopersicum* | 2 |
| | M-0182906 | Germany | 1877 | *Solanum tuberosum* | 2 |
| | M-0182896 | Germany | 1877 | *Solanum tuberosum* | 2 |
| | M-0182904 | Austria | 1879 | *Solanum tuberosum* | 2 |
| | M-0182903 | Canada | 1896 | *Solanum tuberosum* | 2 |
| | KM177512 | England | NA | *Solanum tuberosum* | 1 |
| **Modern samples** | 06_3928A | England | 2006 | *Solanum tuberosum* | 3 |
| | DDR7602 | Germany | 1976 | *Solanum tuberosum* | 4 |
| | P1362 | Mexico | 1979 | *Solanum tuberosum* | 5 |
| | P6096 | Peru | 1984 | *Solanum tuberosum* | 5 |
| | P7722 (*P. mirabilis*) | USA | 1992 | *Solanum lycopersicum* | 5 |
| | P9464 | USA | 1996 | *Solanum tuberosum* | 5 |
| | P12204 | Scotland | 1996 | *Solanum tuberosum* | 5 |
| | P13527 | Ecuador | 2002 | *Solanum andreanum* | 5 |
| | P10127 | USA | 2002 | *Solanum lycopersicum* | 5 |
| | P13626 | Ecuador | 2003 | *Solanum tuberosum* | 5 |
| | P10650 | Mexico | 2004 | *Solanum tuberosum* | 5 |
| | LBUS5 | South Africa | 2005 | *Petunia hybrida* | 6 |
| | P11633 | Hungary | 2005 | *Solanum lycopersicum* | 5 |
| | NL07434 | Netherlands | 2007 | *Solanum tuberosum* | 3 |
| | P17777 | USA | 2009 | *Solanum lycopersicum* | 5 |
| | P17721 | USA | 2009 | *Solanum tuberosum* | 5 |

*1, Kew Royal Botanical Gardens; 2, Botanische Staatssammlung München; 3, (Cooke et al., 2012); 4, (Kamoun et al., 1999); 5, World Oomycete Genetic Resource Collection at UC Riverside, CA; 6, Dr. Adele McLeod, Univ. of Stellenbosch, South Africa

842    **Table 2.** Sequencing strategy.

| ID | Instrument and read type | Sequencing center | Coverage |
|---|---|---|---|
| M-0182896 | HiSeq 2000 (2x101 bp) | MPI | High |
| M-0182897 | HiSeq 2000 (2x101 bp) | MPI | Low[1] |
| M-0182898 | HiSeq 2000 (2x101 bp) | MPI | Low |
| M-0182900 | HiSeq 2000 (2x101 bp) | MPI | Low[2] |
| M-0182903 | HiSeq 2000 (2x101 bp) | MPI | Low |
| M-0182904 | HiSeq 2000 (2x101 bp) | MPI | Low[1] |
| M-0182906 | HiSeq 2000 (2x101 bp) | MPI | Low[2] |
| M-0182907 | HiSeq 2000 (2x101 bp) | MPI | Low |
| KM177497 | MiSeq (2x150 bp) | MPI | Low |
| KM177500 | MiSeq (2x150 bp) | MPI | Low[1] |
| KM177502A | MiSeq (2x150 bp) | MPI | Low[1] |
| KM177507 | MiSeq (2x150 bp) | MPI | Low[1] |
| KM177509 | MiSeq (2x150 bp) & HiSeq 2000 (2x101 bp) | MPI | Low |
| KM177512 | MiSeq (2x150 bp) & HiSeq 2000 (2x101 bp) | MPI | Low |
| KM177513 | MiSeq (2x150 bp) & HiSeq 2000 (2x101 bp) | MPI | Low |
| KM177514 | MiSeq (2x150 bp) & HiSeq 2000 (2x101 bp) | MPI | Low |
| KM177517 | MiSeq (2x150 bp) & HiSeq 2000 (2x101 bp) | MPI | Low |
| KM177548 | MiSeq (2x150 bp) & HiSeq 2000 (2x101 bp) | MPI | Low |
| 06_3928A | GAIIX (2x76 bp) | TSL | High |
| DDR7602 | GAIIX (2x76 bp) | TSL | High |
| LBUS5 | GAIIX (2x76 bp) | TSL | High |
| NL07434 | GAIIX (2x76 bp) | TSL | High |
| P10127 | HiSeq 2000 (2x101 bp) | MPI | Low |
| P10650 | HiSeq 2000 (2x101 bp) | MPI | Low |
| P12204 | HiSeq 2000 (2x101 bp) | MPI | Low |
| P13527 | GAIIX (2x76 bp) | TSL | High |
| P1362 | HiSeq 2000 (2x101 bp) | MPI | Low |
| P13626 | GAIIX (2x76 bp) | TSL | High |
| P11633 | HiSeq 2000 (2x101 bp) | MPI | Low |
| P17721 | HiSeq 2000 (2x101 bp) | MPI | Low |
| P17777 | GAIIX (2x76 bp) | TSL | High |
| P6096 | HiSeq 2000 (2x101 bp) | MPI | Low |
| P7722 | HiSeq 2000 (2x101 bp) | MPI | Low |
| P9464 | HiSeq 2000 (2x101 bp) | MPI | Low[1] |
| PIC99114 | GAIIX (2x76 bp) | TSL | High |
| PIC99167 | GAIIX (2x76 bp) | TSL | High |

843

844    [1]Samples not included in any analysis due to extremely low coverage

845    [2]Samples used only in mtDNA analysis.

846 **Table 3.** Inferred time to most recent common ancestor (TMRCA) for different splits in the mtDNA tree.

| Node | TMRCA (ya) | | |
|------|------------|---|---|
| | Best estimate | Lower 2.5% | Upper 2.5% |
| I/HERB-1, II | 460 | 300 | 643 |
| Ia/Ib, HERB-1 | 234 | 187 | 290 |
| HERB-1 strains | 182 | 168 | 201 |
| IIa, IIb | 142 | 78 | 214 |

847

848

849

850

851

852

853

854

855

856

857

858 **Table 4.** Presence or absence of avirulence effector genes in historic and modern samples, expressed as percentages of effector genes covered by

859 reads. Sequences and polymorphisms are shown in Table 5 and Table 5 – Source data 1.

| *Avr* gene | *R* gene | HERB-1[†] | US-1 | 20th century non-US-1 | | | | | | Outgroups | |
|---|---|---|---|---|---|---|---|---|---|---|---|
| | | | | EC3527 | EC3626 | P17777 | 06_3928A | NL07434 | Merged | Pm PIC99114 | Pip PIC99167 |
| *Avr1* | *R1* | 100 | 100 | 100 | 0 | 100 | 0 | 0 | 100 | 98 | 100 |
| *Avr2* | *R2* | 100 | 100 | 100 | 100 | 100 | 81 | 100 | 77 | 97 | 100 |
| *Avr3a* | *R3a* | 100 | 100 | 100 | 100 | 100 | 100 | 100 | 100 | 0 | 28 |
| *Avr3b* | *R3b* | 0 | 0 | 0 | 0 | 100 | 0 | 0 | 100 | 100 | 100 |
| *Avr4* | *R4* | 100 | 100 | 100 | 100 | 95 | 89 | 100 | 99 | 85 | 92 |
| *Avrblb1* | *Rpi-blb1* | 100 | 100 | 100 | 100 | 100 | 100 | 100 | 100 | 0 | 0 |
| *Avrblb2* | *Rpi-blb2* | 100 | 100 | 100 | 100 | 92 | 100 | 100 | 89 | 88 | 0 |
| *Avrvnt1* | *Rpi-vnt1* | 100 | 100 | 100 | 100 | 100 | 100 | 100 | 100 | 100 | 100 |
| *AvrSmira1* | *Rpi-Smira1* | 100 | 100 | 100 | 100 | 100 | 100 | 100 | 100 | 97 | 100 |
| *AvrSmira2* | *Rpi-Smira2* | 100 | 100 | 100 | 100 | 100 | 100 | 100 | 100 | 100 | 0 |

860
861 † Same sequences obtained for M-0182896 and merged sequences.
862 * Same sequences obtained for DDR7602 and LBUS5.
863
864

865

866

867 **Table 5**. Amino acid differences in the avirulence effectors AVR1, AVR2, AVR3a and

868 AVR4 encoded by the T30-4 reference genome, HERB-1 and DDR7602 (US-1). IDs in

869 parentheses refer to gene models in reference genome. Full-length sequences of deduced

870 amino acid sequences of HERB-1 AVR1, AVR2, AVR3a and AVR4 are provided in Table5 –

871 Source data 1.

872

| Position | Substitution | | | Note |
|---|---|---|---|---|
| | T30-4 | HERB1 | DDR7602 | |
| AVR1 (PITG_16663) | | | | |
| 80 | T | T | T, S | |
| 142 | I | I, T | T | HERB-1 polymorphisms shared with T30-4 |
| 154 | V | V, A | A | and DDR7602. |
| 185 | I | I | I, V | |
| AVR2 (PITG_22870) | | | | |
| 31 | N | K | K | HERB-1 identical to DDR7602. |
| AVR3a (PITG_14371) | | | | |
| 19 | S | C | C | |
| 80 | E | K | K | HERB-1 identical to DDR7602; both |
| 103 | M | I | I | correspond to AVR3a$^{Kl}$ isoform. |
| 139 | M | L | L | |
| AVR4 (PITG_07387) | | | | |
| 19 | T | T, I | T | |
| 139 | L | S | L, S | HERB-1 polymorphisms shared with T30-4 |
| 221 | L | V | L, V | and DDR7602. |
| 271 | V | F | V, F | |

873
874
875

876 **Table 5 – Source data 1.** Full-length sequences of deduced amino acid sequences of HERB-1
877 AVR1, AVR2, AVR3a and AVR4.

878

879

880 **Figure 1. Countries of origin of samples used in whole-genome, mtDNA genome or both**

881 **analyses.** Red indicates number of historic and blue of modern samples. More information on

882 the samples is given in Table 1 and Table 2.

883

884 **Figure 2. Ancient DNA-like characteristic of historic samples. (A)** Lengths of merged

885 reads from historic sample M-0182898. **(B)** Mean lengths of merged reads from historic

886 samples. **(C)** Nucleotide mis-incorporation in reads from the historic sample M-0182898. **(D)**

887 Deamination at first 5' end base in historic samples. **(E)** Percentage of merged reads that

888 mapped to the *P. infestans* reference genome.

889

**Figure 3. Coverage and SNP statistics. (A)** Mean nuclear genome coverage from historic (red) and modern (blue) samples. **(B)** Homo- and heterozygous SNPs in each sample. **(C)** Inverse cumulative coverage for all homozygous SNPs across all samples. **(D)** Same as (C) for homo- and heterozygous SNPs.

**Figure 3 – figure supplement 1. Accuracy and sensitivity of SNP calling at different cutoffs for SNP concordance based on 3- and 50-fold coverage of simulated data.** Rescue cov. – minimum coverage required to accept SNP calls in low-coverage genomes based on these SNPs having been found in high-coverage genomes. The cutoffs enclosed in orange rectangles were used for the final analysis.

**Figure 4. Maximum-parsimony phylogenetic tree of complete mtDNA genomes.** Sites with less than 90% information were not considered, leaving 24,560 sites in the final dataset. Numbers at branches indicate bootstrap support (100 replicates), and scale indicates changes.

**Figure 4 – figure supplement 1. Maximum-likelihood phylogenetic tree of complete mtDNA genomes.** Sites with less than 90% information were not considered, leaving 24,560 sites in the final dataset. Numbers at branches indicate bootstrap support (100 replicates).

**Figure 4 – figure supplement 2. mtDNA sequences around diagnostic *Msp1* restriction site (grey) for reference haplotype modern strains (blue) and historic strains (red).** The *Msp1* (CCGG) restriction site is only present in the Ib haplotype; all other strains have a C-to-T substitution (CTGG).

**Figure 5. Correlation between nucleotide distance of mtDNA genomes of HERB-1/haplotype Ia/haplotype Ib clade to the outgroup P17777 and sample age in calendar years before present.**

**Figure 6. Divergence estimates of mtDNA genomes.** Bayesian consensus tree from 147,000 inferred trees. Posterior probability support above 50% is shown next to each node. Blue horizontal bars represent the 95% HPD interval for the node height. Light yellow bars indicate major historical events discussed in the text. See Figure 5 – Table 3 for detailed estimates at the four main nodes in *P. infestans*.

924 **Figure 7. Phylogenetic trees of high-coverage nuclear genomes using both homozygous**

925 **and heterozygous SNPs. (A)** Maximum-parsimony tree, considering only sites with at least

926 95% information, leaving 4,498,351 sites in the final dataset. Numbers at branches indicate

927 bootstrap support (100 replicates), and scale indicates genetic distance. **(B)** Maximum-

928 likelihood tree. **(C)** Heat map of genetic differentiation (color scale indicates SNP

929 differences). US-1 strains DDR7062 and LBUS5 have the genomes sequences closest to M-

930 0182896 (asterisks). The two US-1 isolates in turn are outliers compared to all other modern

931 strains (highlighted by a gray box).

932

933 **Figure 7 – figure supplement 1. Phylogenetic trees of high- and low-coverage nuclear**

934 **genomes. (A)** Neighbor-joining tree of high-coverage genomes using 4,595,012 homo- and

935 heterozygous SNPs. Numbers at branches indicate bootstrap support (100 replicates), and

936 scale indicates genetic distance. **(B)** Neighbor-joining tree of high- and low-coverage

937 genomes using 2,101,039 homozygous and heterozygous SNPs. Numbers at branches indicate

938 bootstrap support above 50, from 100 replicates) Scale indicates genetic distance. **(C)**

939 Maximum parsimony tree of high- and low-coverage genomes using 315,394 SNPs

940 homozygous and heterozygous SNPs (using only sites with at least 80% information).

941

942 **Figure 8. Ploidy analysis. (A)** Diagram of expected read frequencies of reads at biallelic

943 SNPs for diploid, triploid and tetraploid genomes. **(B)** Reference read frequency at biallelic

944 SNPs in gene dense regions (GDRs) for the historic sample M-0182896, two modern samples,

945 and simulated diploid, triploid and tetraploid genomes. The simulated tetraploid genome is

946 assumed to have 20 % of pattern 1 and 80 % of pattern 3 shown in (A). The shape and

947 kurtosis of the observed distributions are similar to the corresponding simulated ones. **(C)**

948 Polymorphic positions with more than one allele in the GDR.

949

950 **Figure 8 – figure supplement 1. Reference read frequency at biallelic SNPs in gene dense**

951 **regions (GDRs) for five modern high-coverage samples.**

952

953 **Figure 9. Read allele frequencies of historic genome M-0182896 and US-1 isolate**

954 **DDR7602.** Alleles were classified as ancestral or derived using outgroup species *P. mirabilis*

955 and *P. ipomoeae*. There were 40,532 segregating sites. **(A)** Distributions of derived alleles at

956 sites segregating between M-0182896 and DDR7602. **(B)** Annotation of the different site
957 classes.

958

959 **Figure 10. The effector gene *Avr3a* and its cognate resistance gene *R3a*.** (**A**) Diagram of
960 AVR3A effector protein. (**B**) Frequency of *Avr3a* alleles in historic and modern *P. infestans*
961 strains. (**C**) Neighbor-joining tree of *R3a* homologs from potato, based on 0.67 kb partial
962 nucleotide sequences of *S. tuberosum R3a* (blue, accession number AY849382.1) and
963 homologs (dark grey) in GenBank, and *de novo* assembled contigs from M-0182896 (red).
964 Numbers at branches indicate bootstrap support with 500 replicates. Scale indicates changes.

965

966 **Figure 10 – figure supplement 1. Summary of *de novo* assembly of RXLR effector genes.**
967 TBLASTN query was performed with 549 RXLR proteins as a query and contigs as a
968 database. When the High-scoring Segment Pair (HSP) and matched amino acids both covered
969 ≥99% of the query length, we recorded a hit. Results with the optimal *k*-mer size are
970 highlighted.

971

972 **Figure 11. Suggested paths of migration and diversification of *P. infestans* lineages**
973 **HERB-1 and US-1.** The location of the metapopulation that gave rise to HERB-1 and US-1
974 remains uncertain; here it is proposed to have been in North America.

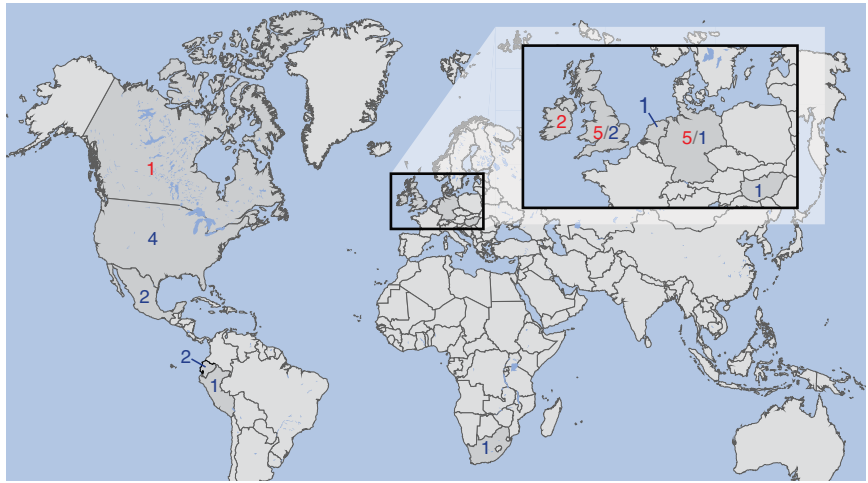

**A**

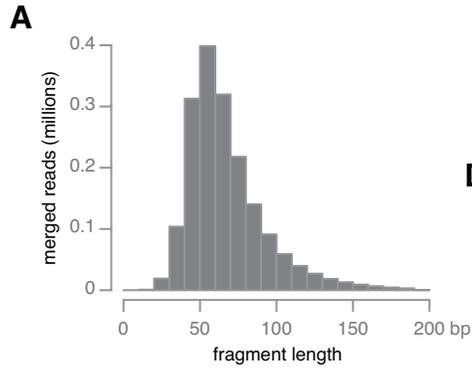

**B**

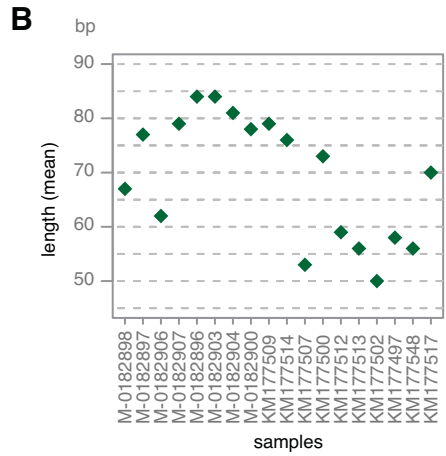

**C**

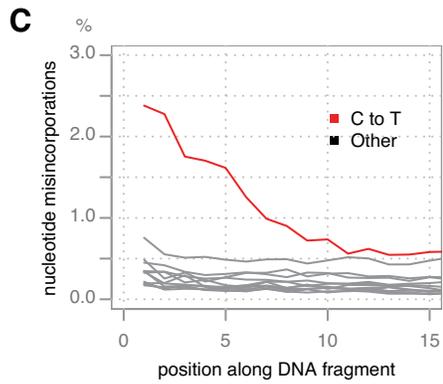

**D**

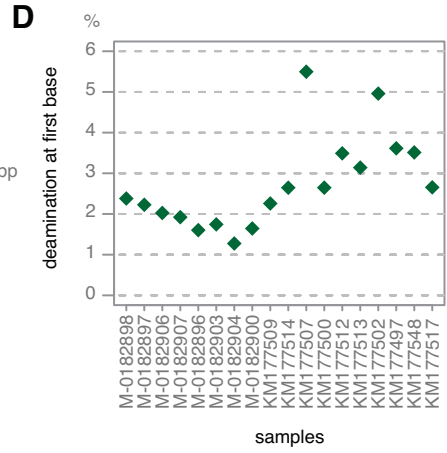

**E**

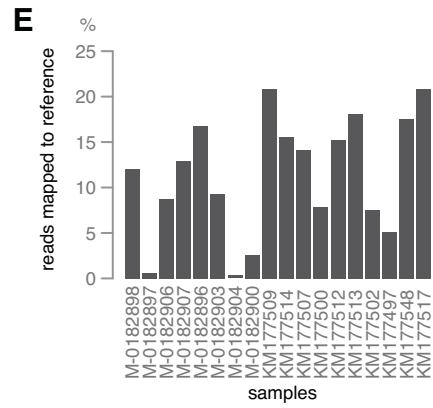

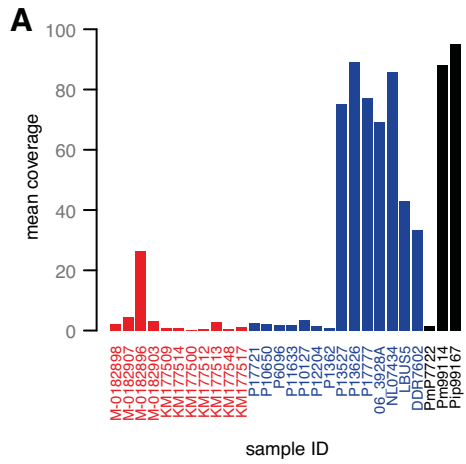
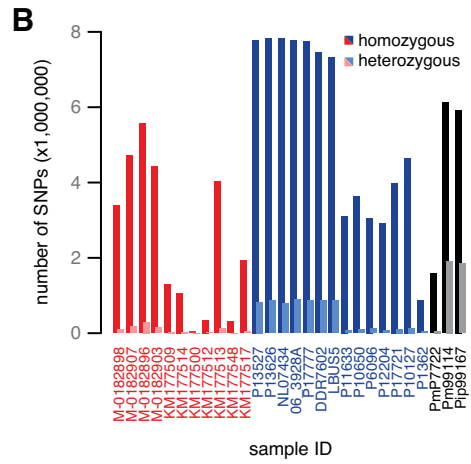
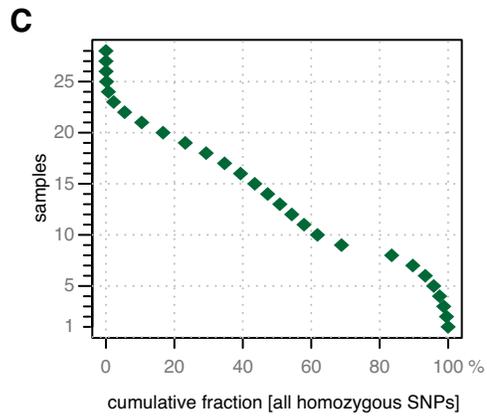
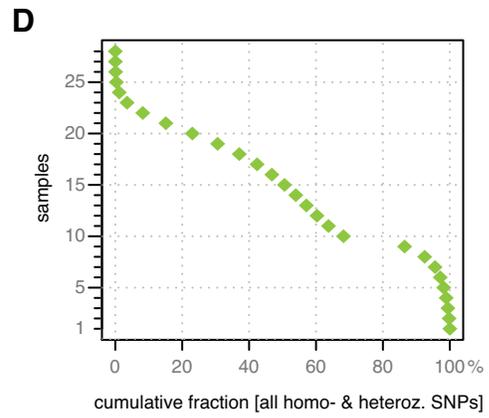

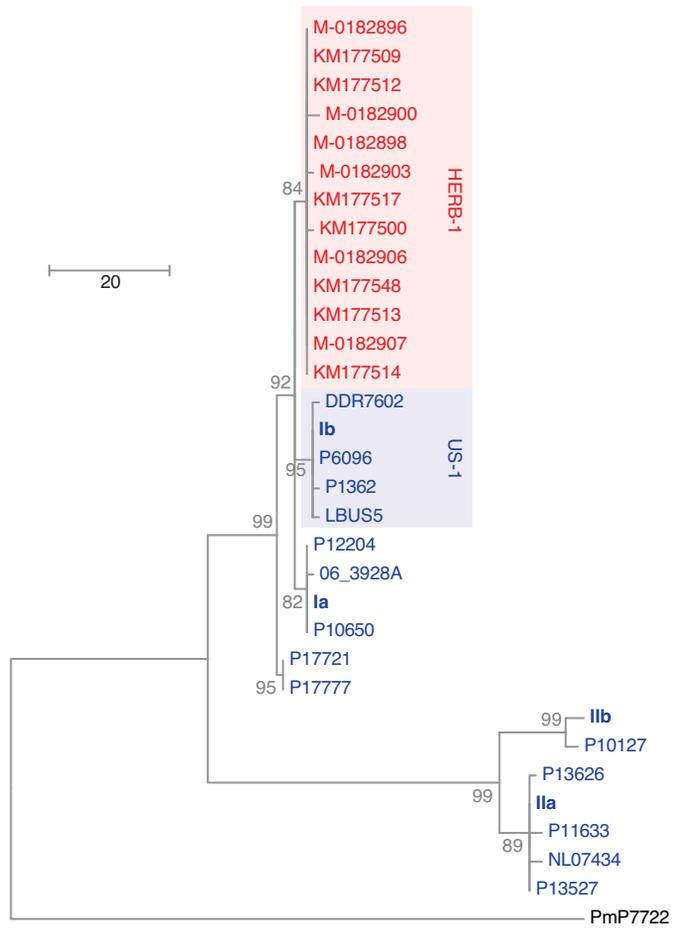

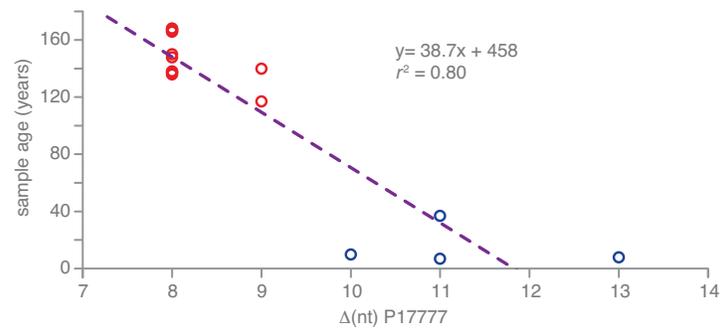

Spanish Conquest

Irish Famine

P17777
LBUS5
**Ib**
DDR7602
06_3928A
**Ia**
KM177500
KM177512
KM177513
M-0182900
M-0182906
M-0182896
KM177517
M-0182903
M-0182907
KM177509
M-0182898
KM177514
KM177548
P13626
**IIa**
NL07434
P13527
**IIb**
PmP7722

70
100
100
94
100
100
100
100
100
100
100
100
100

750    1000    1250    1500    1750    2000 CE

**A**

DDR7602
LBUS5
M-0182896
P13527
P13626
NL07434
06_3928A
P17777
Pip99167
Pm99114

200,000

**B**

DDR7602
LBUS5
M-0182896
P13527
P13626
NL07434
06_3928A
P17777
Pip99167
Pm99114

0.02

**C**

P17777
06_3928A
NL07434
P13626
P13527
LBUS5
DDR7602
M-0182896

M-0182896  DDR7602  LBUS5  P13527  P13626  NL07434  06_3928A  P17777

360k
340k
320k
300k
280k
260k
240k

**A**

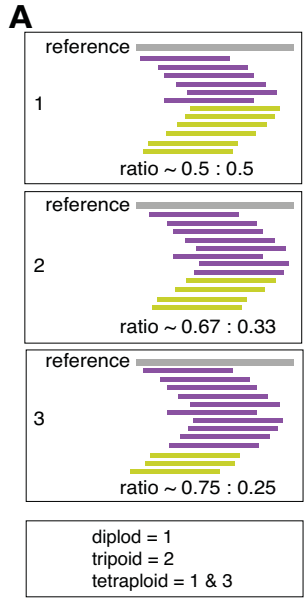

reference
1
ratio ~ 0.5 : 0.5

reference
2
ratio ~ 0.67 : 0.33

reference
3
ratio ~ 0.75 : 0.25

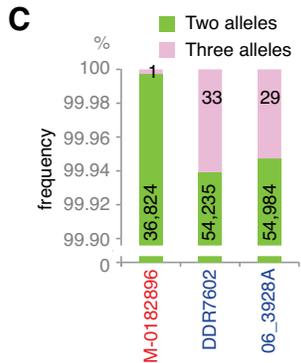

diplod = 1
tripiod = 2
tetraploid = 1 & 3

**B**

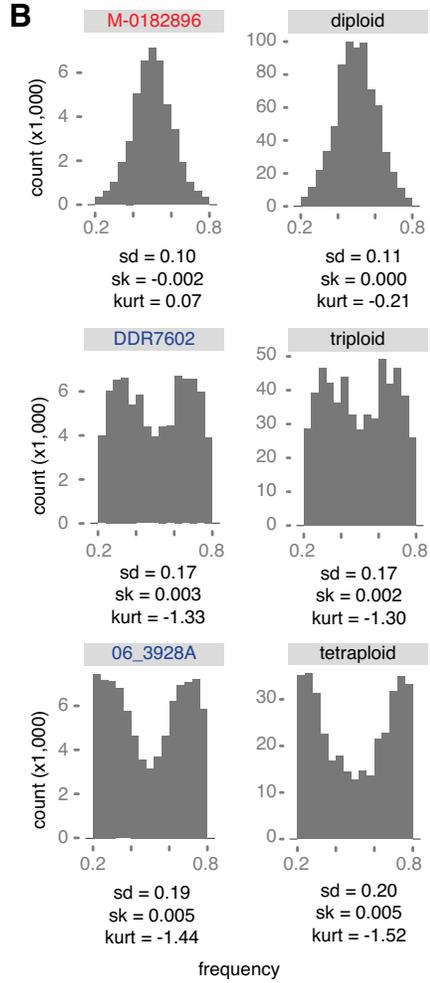

M-0182896

count (x1,000)

0.2    0.8

sd = 0.10
sk = -0.002
kurt = 0.07

diploid

count (x1,000)
100
80
60
40
20

0.2    0.8

sd = 0.11
sk = 0.000
kurt = -0.21

DDR7602

count (x1,000)
6
4
2

0.2    0.8

sd = 0.17
sk = 0.003
kurt = -1.33

triploid

count (x1,000)
50
40
30
20
10
0

0.2    0.8

sd = 0.17
sk = 0.002
kurt = -1.30

06_3928A

count (x1,000)
6
4
2

0.2    0.8

sd = 0.19
sk = 0.005
kurt = -1.44

tetraploid

count (x1,000)
30
20
10

0.2    0.8

sd = 0.20
sk = 0.005
kurt = -1.52

frequency

**C**

%

■ Two alleles  ■ Three alleles

frequency

100
99.98
99.96
99.94
99.92
99.90

1
36.824
M-0182896

33
54.235
DDR7602

29
54.984
06_3928A

**A**

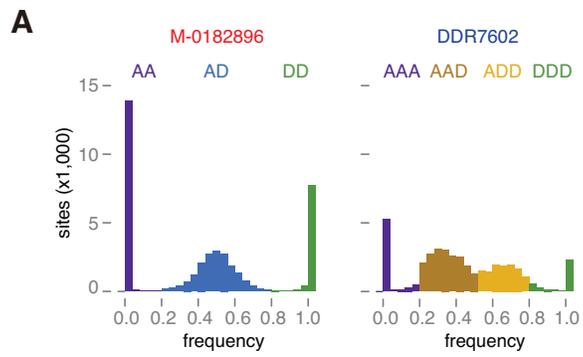

**B**

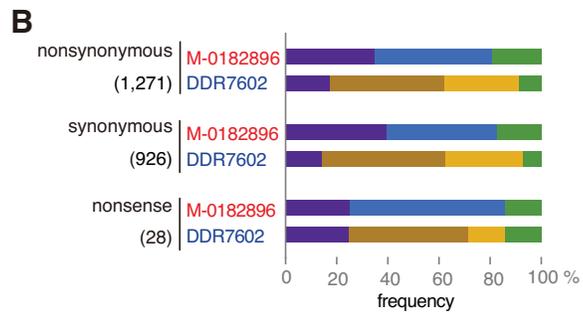

**A**

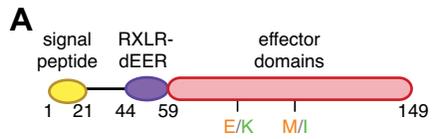

**B**

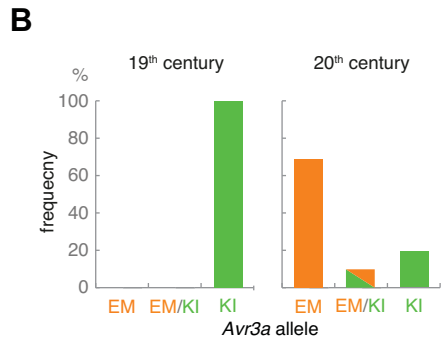

**C**

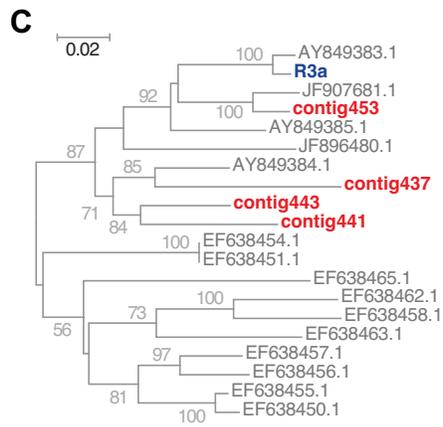

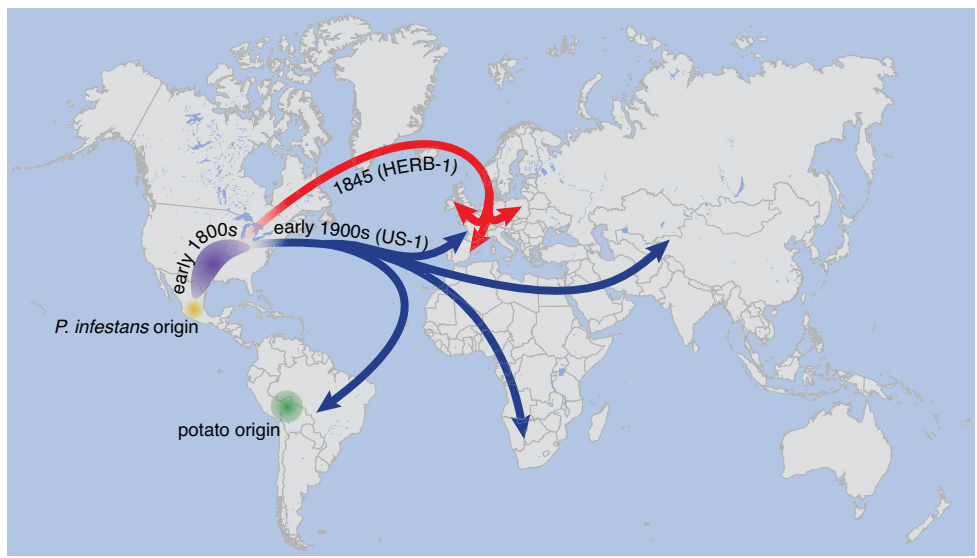

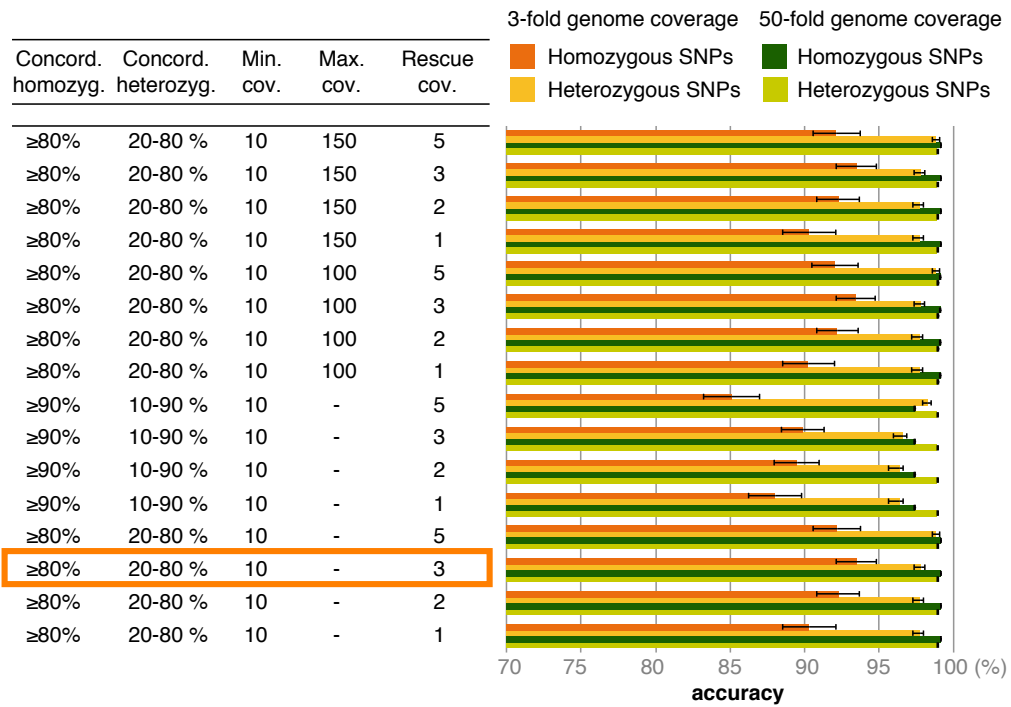
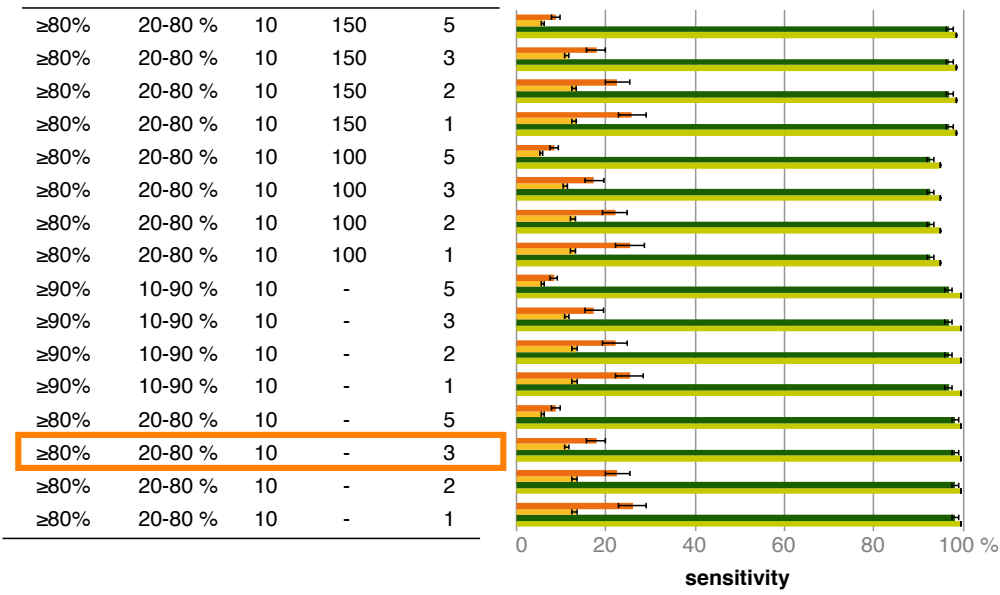

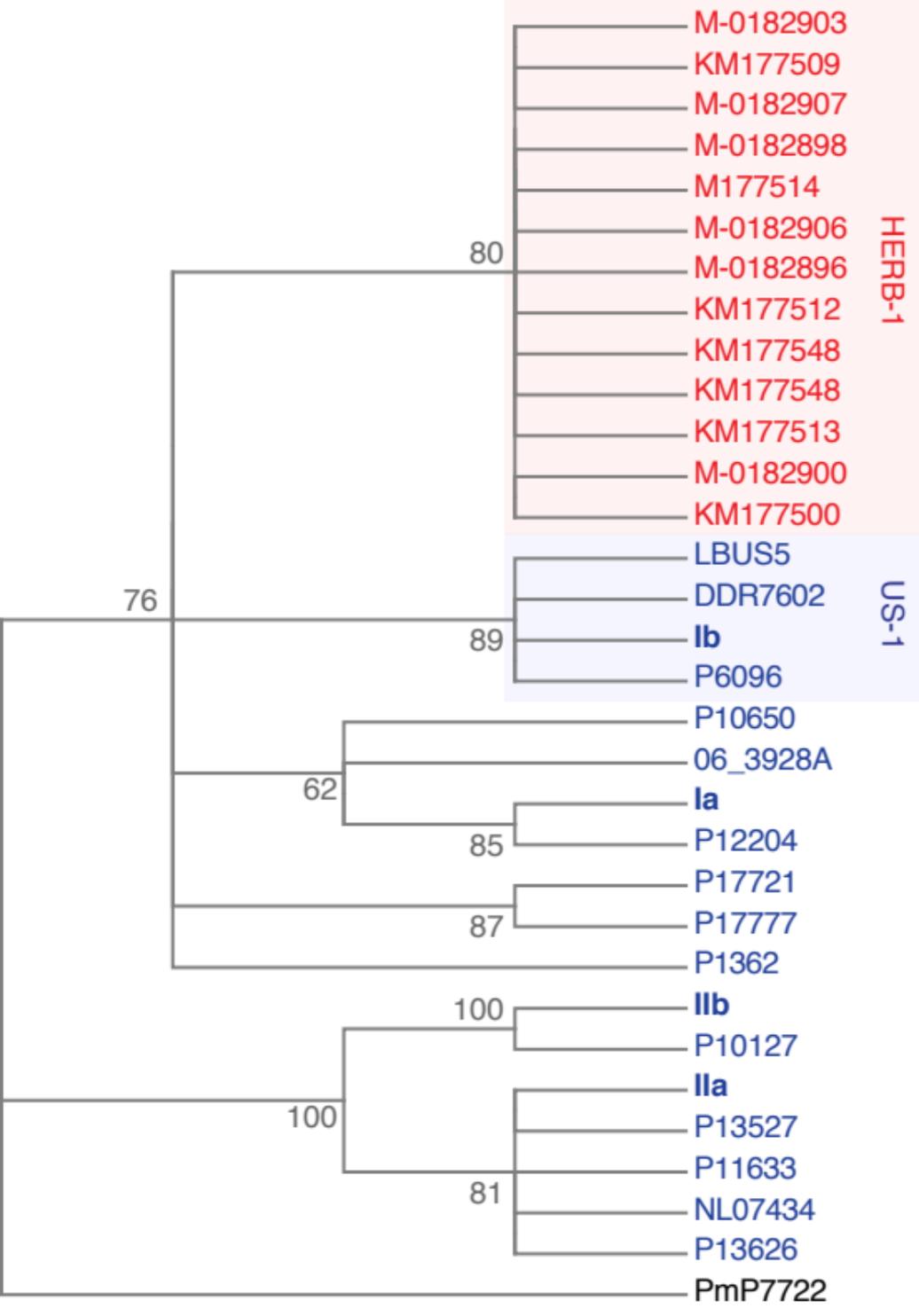

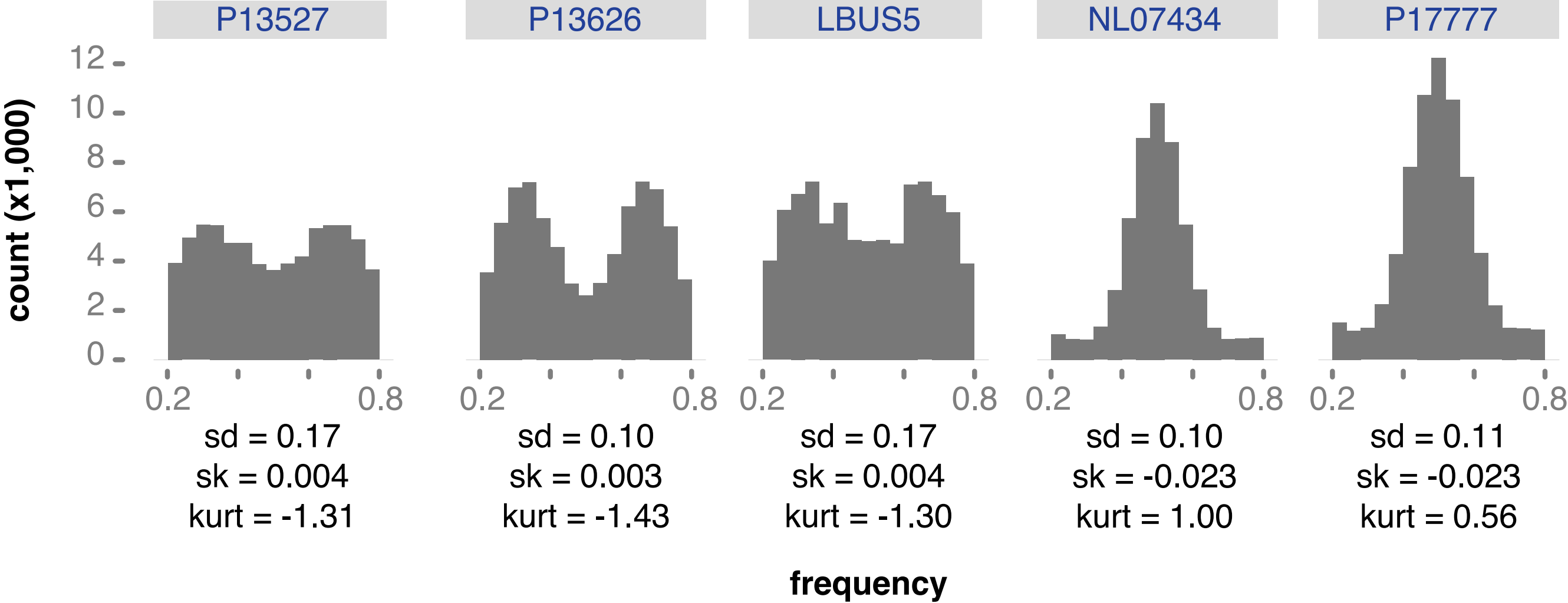

| kmer | N50 (bp) | Longest Contig (bp) | RXLR proteins with TBLASTN hit | AVR1 | AVR2 | AVR3a$^{EM}$ | AVR4 |
|---|---|---|---|---|---|---|---|
| 41 | 160 | 2,686 | 25 | 48 | 81 | 37 | 66 |
| 51 | 169 | 4,616 | 61 | 39 | 100 | 46 | 78 |
| 61 | 191 | 8,487 | 89 | 100 | 100 | 84 | 87 |
| 63 | 219 | 9,428 | 90 | 77 | 100 | 84 | 87 |
| 65 | 241 | 9,430 | 92 | 100 | 99 | 84 | 87 |
| 67 | 248 | 9,509 | 90 | 100 | 99 | 97 | 79 |
| 69 | 257 | 15,874 | 88 | 100 | 99 | 97 | 79 |
| 71 | 259 | 14,583 | 76 | 100 | 91 | 97 | 45 |
| 81 | 312 | 36,586 | 76 | 100 | 99 | 97 | 45 |
| 91 | 362 | 73,532 | 48 | 75 | 79 | 90 | 44 |
| 101 | 393 | 68,999 | 19 | 75 | 79 | 90 | 7 |
| 111 | 429 | 68,999 | 15 | 75 | 65 | 69 | 7 |
| 121 | 473 | 25,791 | 6 | 7 | 8 | 51 | 7 |

Matched aa / AVR length (%)

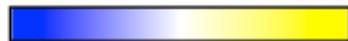

0          50          100